\documentstyle[12pt]{article}
\newcommand\be{\begin{equation}}
\newcommand\ee{\end{equation}}
\newcommand\bea{\begin{eqnarray}}
\newcommand\eea{\end{eqnarray}}
\newcommand\ket[1]{|#1\rangle}

\newcommand\braket[2]{\langle #1|#2\rangle}
\newcommand{\fatalpha}{{\bf \alpha \kern -0.44em \alpha}}
\newcommand{\fatsigma}{{\bf \sigma \kern -0.54em \sigma}}
\newcommand{\tpchi}{{\bf \chi \kern -0.35em \chi}}
\newcommand{\llambda}{{\bf \lambda \kern -0.45em \lambda}}

\renewcommand{\theequation}{\arabic{equation}}
\renewcommand{\theequation}{\thesection-\arabic{equation}}
\bibliography{plain}
\title{\bf {Investigation of continuous-time quantum walk on root lattice $A_n$ and honeycomb lattice}}\vspace{20mm}
\author{ M. A. Jafarizadeh$^{a,b,c}$
\thanks{E-mail:jafarizadeh@tabrizu.ac.ir} ,
R.Sufiani$^{a,b}$
\thanks{E-mail:sofiani@tabrizu.ac.ir}
\\ $^a${\small Department of Theoretical Physics and Astrophysics,
Tabriz University, Tabriz 51664, Iran.} \\ $^b${\small Institute
for Studies in Theoretical Physics and Mathematics, Tehran
19395-1795, Iran.} \\ $^c${\small Research Institute for
Fundamental Sciences, Tabriz 51664, Iran.}} \pagebreak

\vspace{20mm}
\begin{document}
\maketitle \vspace{15mm}
\newpage
\begin{abstract}
The continuous-time quantum walk (CTQW) on root lattice $A_n$ (
known as hexagonal lattice for $n=2$) and honeycomb one is
investigated by using spectral distribution method. To this aim,
some association schemes are constructed from abelian group
$Z^{\otimes n}_m$ and two copies of finite hexagonal lattices,
such that their underlying graphs tend to root lattice $A_n$ and
honeycomb one, as the size of the underlying graphs grows to
infinity. The CTQW on these underlying graphs is investigated by
using the spectral distribution method and stratification of the
graphs based on Terwilliger algebra, where we get the required
results for root lattice $A_n$ and honeycomb one, from large
enough underlying graphs. Moreover, by using the stationary phase
method, the long time behavior of  CTQW on infinite graphs is
approximated with finite ones. Also it is shown that the
Bose-Mesner algebras of our constructed association schemes
(called $n$-variable $P$-polynomial) can be generated by $n$
commuting generators, where raising, flat and lowering operators
(as elements of Terwilliger algebra) are associated with each
generator. A system of $n$-variable orthogonal polynomials which
are special cases of \textit{generalized} Gegenbauer polynomials
is constructed, where the probability amplitudes are given by
integrals over these polynomials or their linear combinations.
Finally the suppersymmetric structure of finite honeycomb lattices
is revealed.


{\bf PACs Index: 03.65.Ud }
\end{abstract}
\vspace{70mm}
\newpage
\section{Introduction}
 Quantum walks have recently been introduced and
investigated with the hope that they may be useful in constructing
new efficient quantum algorithms (for reviews of quantum walks,
see \cite{Amb.}, \cite{Kemp.}, \cite{Treg.}). A study of random
walks on simple lattices is well known in physics(see
\cite{Feyn.}). Recent studies of quantum walks on more general
graphs were described in \cite{fg}, \cite{fg1}, \cite{Amb.},
\cite{Aharanov}, \cite{Childs}. Some of these works study the
problem in the important context of algorithmic problems on graphs
and suggest that quantum walks is a promising algorithmic
technique for designing future quantum algorithms.

On the other hand, the theory of association schemes
\cite{Ass.sch.} (the term of association scheme was first coined
by R. C. Bose and T. Shimamoto in \cite{Bose}) has its
  origin in the design of statistical
experiments. The connection of association schemes to algebraic
codes, strongly regular graphs, distance regular graphs, design
theory etc., further intensified their study. A further step in
the study of association schemes was their algebraization. This
formulation was done by R. C. Bose and D. M. Mesner who introduced
an algebra generated by the adjacency matrices of the association
scheme, known as Bose-Mesner algebra. The other formulation was
done by P. Terwilliger, known as the Terwilliger algebra. This
algebra has been used to study $P$- and $Q$-polynomial schemes
\cite{T1}, group schemes \cite{jmbmo,ebam}, and Doob schemes
\cite{kta}.

Authors in \cite{js,js1,jss} have introduced a new method for
calculating the probability amplitudes of CTQW on particular
graphs based on spectral distribution and algebraic combinatorics
structures of the graphs, where a canonical relation between the
interacting Fock space of CTQW (i.e., Hilbert space of CTQW
starting from a given site which consists of irreducible submodule
of Terwilliger algebra with maximal dimension) and a system of
orthogonal polynomials has been established which leads to the
notion of quantum decomposition (QD) introduced in
\cite{nob,obah}.  In \cite{js,js1,obah}, only the particular
graphs of QD type have been studied, where the adjacency matrices
posses quantum decomposition and one can give the graph a
three-term recursion structure. Then, by employing the three-term
recursion structure of the graph, one can define the Stieltjes
transform of spectral distribution and obtain the corresponding
spectral distribution via inverse Stieltjes transform. The QD
property is inherent in underlying graphs of $P$-polynomial
association schemes (for more details of $P$-polynomial
association schemes, see \cite{bannai}, \cite{Dels}, \cite{Leo},
\cite{T1}, \cite{T2}) due to the algebraic combinatorics structure
of schemes, particularly the existence of raising, flat and
lowering operators.

Here in this work, we investigate CTQW on root lattice $A_n$ and
honeycomb one by using spectral distribution method. In particular,
we discuss the root lattice $A_2$ (called hexagonal or triangular
lattice) in more details, and then generalize the results to the
case of $A_n$. To this purpose, first we construct some interesting
association schemes from abelian group $Z^{\otimes n}_m$ and finite
honeycomb lattice, where in the first case, the orbits of Weyl group
corresponding to the finite lattice, define a translation invariant
(non-symmetric) association scheme on $Z^{\otimes n}_m$. Then, by
symmetrization method, we construct a new symmetric association
scheme, where CTQW is investigated on its underlying graph. In the
latter case , we construct the association scheme from two copies of
finite hexagonal lattices, where the corresponding adjacency matrix
$A$ is defined suitably from the adjacency matrix of finite
hexagonal lattice and the other adjacency matrices are constructed
via powers of $A$ (in this case we have not a systematic procedure
for construction of association scheme as in the first case). These
association schemes have the privileges that, for large enough size
of their underlying graphs, they tend to root lattice $A_n$ and
honeycomb one, respectively. By using spectral distribution method,
we study CTQW on these underlying graphs via their algebraic
combinatorics structures such as (reference state dependent)
Terwilliger algebras. By choosing the starting site of the walk as
reference state, the Terwilliger algebra connected with this choice,
stratifies the graph  into disjoint unions of strata, where the
amplitudes of observing the CTQW on all sites belonging to a given
stratum are the same. This stratification is different from the one
based on distance, i.e., it is possible that two strata with the
same distance from starting site posses different probability
amplitudes. Then we study the CTQW on root lattice $A_n$ and
honeycomb one by using the results of finite lattices. Moreover, by
using the stationary phase method \cite{stat.book}, the long time
behavior of the quantum walk on infinite graphs is approximated with
finite ones. In fact, the numerical results show that, the $A_2$
(honeycomb) lattice  can be approximated by a finite hexagonal
(finite honeycomb) lattice for $m$ larger than $\sim 50$ ($\sim 60$)
and times $t\sim 1000$ ($t\sim 700$).

Another interesting property of constructed association schemes from
$Z^{\otimes n}_m$ is that, their corresponding Bose-Mesner algebras
are generated by $n$ commuting generators. In particular, the
adjacency matrices are $n$-variable polynomials of the generators,
where recursion relations for the polynomials are given by using the
structure of the association schemes. This property allows us to
generalize the notion of $P$-polynomial association schemes to
\textit{$n$-variable} $P$-polynomial association schemes, where the
spectral distributions associated with the generators are functions
of $n$ variables (variables assigned to the generators). Also, we
associate raising, lowering and flat operators with each generator
via the elements of corresponding Terwilliger algebra. Then, by
using the recursion relations associated with the Bose-Mesner
algebra, we construct a system of $n$-variable orthogonal
polynomials which are special cases of orthogonal polynomials known
as \textit{generalized} Gegenbauer polynomials \cite{Per1},
\cite{Per2}, where the probability amplitudes of the walk are given
by integrals over these polynomials or their linear combinations. In
fact, it is shown that similar to the $P$-polynomial case, there is
a canonical isomorphism from the interacting Fock space of CTQW on
finite root lattice $A_n$ onto the closed linear span of these
orthogonal polynomials. Finally, we reveal the suppersymmetric
structure of finite honeycomb lattices in the appendix.

The organization of the paper is as follows. In section $2$, we
introduce briefly root lattice $A_n$ and honeycomb lattice. In
section 3, we give a brief outline of association schemes,
Bose-Mesner and Terwilliger algebras. In section $4$, we give an
algorithm for constructing some underlying graphs of so called
two-variable $P$-polynomial association schemes and then following
ref.\cite{js1}, we stratify the underlying graphs of constructed
two-variable $P$-polynomial association schemes. In section $5$, we
give a brief review of spectral distribution method and discuss the
construction of two-variable orthogonal polynomials. Section $6$, is
devoted to CTQW on hexagonal lattice and honeycomb one, by using
spectral distribution method. Also, the asymptotic behavior of
probability amplitudes of the walk at large time $t$, is discussed.
Finally,  we generalize the discussions of $A_2$ to $A_n$ in section
$7$. The paper is ended with a brief conclusion together with an
appendix on the suppersymmetric structure of finite honeycomb
lattices.

\section{Root lattice $A_n$ and honeycomb lattice}
\subsection{Root lattice $A_n$}
It is well known that a Coxeter-Dynkin diagram determines a system
of simple roots in the Euclidean space $E_n$. The finite group
$W$, generated by the reflections through the hyperplanes
perpendicular to roots $\alpha_i$, $i=1,...,n$
\begin{equation}\label{wyle}
r_{i}(\beta)=\beta-2\frac{(\alpha_i,\beta)}{(\alpha_i,\alpha_i)}\alpha_i\in
R,
\end{equation}
is called a Weyl group (for the theory of such groups, see
\cite{wyle} and \cite{wyle1}). An action of elements of the Wyle
group $W$ upon simple roots leads to a finite system of vectors,
which is invariant with respect to $W$. A set of all these vectors
is called a system of roots associated with a given Coxeter-Dynkin
diagram (for a description of the correspondence between simple
Lie algebras and Coxeter-Dynkin diagrams, see, for example,
\cite{coxter}). It is proven that roots of $R$ are linear
combinations of simple roots with integral coefficients. Moreover,
there exist no roots which are linear combinations of simple roots
$\alpha_i$, $i=1,2,...,n$, both with positive and negative
coefficients. The set of all linear combinations
\begin{equation}
Q=\{\sum_{i=1}^n a_i\alpha_i\;\ |\;\ a_i \in Z\}\equiv \bigoplus_i
Z\alpha_i,
\end{equation}
 is called a root lattice corresponding
to a given Coxeter-Dynkin diagram. Root system $R$ which
corresponds to Coxeter-Dynkin diagram of Lie algebra of the group
$SU(n+1)$, gives root lattice $A_n$. For example root system $A_2$
(corresponding to lie algebra of SU(3)) is shown in Fig.$1$, where
the roots form
 a regular hexagon and $\alpha$ and $\beta$ are simple roots (see
 Fig.1). This lattice sometimes is called hexagonal lattice or
 triangular lattice.

It is convenient to describe root lattice, Weyl group and its
orbits for the case of $A_n$ in the subspace of the Euclidean
space $E_{n+1}$ , given by the equation
\begin{equation}
x_1+x_2+...+x_{n+1}=0,
\end{equation}
where $x_1$, $x_2$,..., $x_{n+1}$ are the orthogonal coordinates
of a point $x\in E_{n+1}$. The unit vectors in directions of these
coordinates are denoted by $e_j$, respectively. Clearly, $e_i\perp
e_j$, $i\neq j$. The set of roots is given by the vectors
\begin{equation}
\alpha_{ij}=e_i-e_j,\;\;\;\ i\neq j.
\end{equation}
The roots $\alpha_{ij}$, with $i<j$ are positive and the roots
\begin{equation}
\alpha_i\equiv \alpha_{i,i+1}=e_i-e_{i+1},\;\;\;\ i=1,...,n,
\end{equation}
constitute the system of simple roots.

By means of the formula (\ref{wyle}), one can find that the
reflection $r_{{\alpha}_{ij}}$ acts upon the vector
$\lambda=\sum_{i=1}^{n+1} m_ie_i$, given by orthogonal
coordinates, by permuting the coordinates $m_i$ and $m_j$. Thus,
$W(A_n)$ (Weyl group corresponding to $A_n$) consists of all
permutations of the orthogonal coordinates $m_1$, $m_2$,...,
$m_{n+1}$ of a point $\lambda$, that is, $W(A_n)$ coincides with
the symmetric group $S_{n+1}$. The orbit $O(\lambda)$,
$\lambda=(m_1,m_2,...,m_{n+1})$, consists of all different points
$(m_{i_1},m_{i_2},...,m_{i_{n+1}})$ obtained from
$(m_1,m_2,...,m_{n+1})$ by permutations.

For our purposes in this paper, we will construct an underlying
graph of association scheme from abelian group $Z^{\otimes n}_m$
($m\geq3$), such that the constructed graph can be viewed as root
lattice $A_n$ where $Z$ is replaced with $Z_m$.
 \subsection{Honeycomb lattice}
The honeycomb lattice is defined by two sets of direction vectors
(vectors with integer components), but first we should introduce
the notion of odd and even vertices. A vertex is odd if the sum of
its components is odd, otherwise it is even. The honeycomb lattice
is a two dimensional lattice defined as follows\\
\textbf{Definition 4} For an even vertex, the set of direction
vectors is $\{(1,0),(-1,0),(0,1)\}$ and for an odd vertex, the set
of direction vectors is $\{(1,0),(-1,0),(0,-1)\}$.

A honeycomb structure is related to a hexagonal lattice in the
following two ways \\ (1) The centers of the hexagons of a
honeycomb form a hexagonal
lattice, with the rows oriented the same.\\
(2) The vertices of a honeycomb, together with their centers, form
a hexagonal lattice, rotated by the angle of $\pi/6$, and scaled
by a factor $1/\sqrt{3}$, relative to the other lattice.

The ratio of the number of vertices and the number of hexagons is
$2$ (see Fig.2).

In section $4$, we will construct an underlying graph of
association scheme from two copies of hexagonal lattices, where
the graph is equivalent to honeycomb lattice as the size of the
graph grows to infinity.

 \section{Association schemes and their Terwilliger algebra}
In this section, we give a brief review of some of the main
features of symmetric association schemes. For further information
about association schemes, the reader is referred to
\cite{Ass.sch.}, \cite{Bose}, \cite{T1}.

\textbf{Definition 3}\label{def1} (Symmetric association schemes).
Let $V$ be a set of vertices, and $R_i$ $(i = 0,...,d)$ be
nonempty relations on $V$. If the following conditions (1), (2),
(3), and (4) be satisfied, then the pair $Y=(V, \{R_i\}_{0\leq
i\leq d})$ consisting of a vertex set $V$ and a set of relations
$\{R_i\}_{0\leq i\leq d}$ is called an association
scheme.\\
$(1)\;\ \{R_i\}_{0\leq i\leq d}$ is a partition of $V\times V$\\
$(2)\;\ R_0=\{(\alpha, \alpha) : \alpha\in V \}$\\
$(3)\;\ R_i=R_i^t$ for $0\leq i\leq d$, where
$R_i^t=\{(\beta,\alpha) :
(\alpha, \beta)\in R_i\} $\\
$(4)$ Given $(\alpha, \beta)\in R_k$, $p_{ij}^{k}=\mid \{\gamma\in
V : (\alpha, \beta)\in R_i \;\ and \;\ (\gamma,\beta)\in
R_j\}\mid$, where the constants $p_{ij}^{k}$ are called the
intersection numbers, depend only on $i, j$ and $k$ and not on the
choice of $(\alpha, \beta)\in R_k$.

The underlying graph $\Gamma=(V,R_1)$ of an association scheme
 is an undirected connected graph, where the set
$V$ and $R_1$ consist of its vertices and edges, respectively.
Obviously replacing $R_1$ with one of the other relations such as
$R_i$, for  $i\neq 0,1$ will also give us an underlying graph
$\Gamma=(V,R_i)$ (not necessarily a connected graph ) with the
same set of vertices but a new set of edges $R_i$.

 Let $C$ denote the field of complex numbers. By $Mat_V(C)$ we mean the set of all
$n\times n$ matrices over $C$ whose rows and columns are indexed
by $V$. For each integer $i$ ($0\leq i \leq d$), let $A_i$ denote
the matrix in $Mat_V (C)$ with $(\alpha,\beta)$-entry as
\begin{equation}\label{adj.}
\bigl(A_{i})_{\alpha, \beta}\;=\;\cases{1 & if $\;(\alpha,
\beta)\in R_i$,\cr 0 & otherwise\cr}\qquad \qquad (\alpha, \beta
\in V).
\end{equation}
The matrix $A_i$ is called an adjacency matrix of the association
scheme. We then have $ A_0=I$ (by (2) above)  and
\begin{equation}\label{li}
A_iA_j=\sum_{k=0}^{d}p_{ij}^kA_{k},
\end{equation}
so $A_0, A_1, ..., A_d$ form a basis for a commutative algebra
\textsf{A} of $Mat_V(C)$, where \textsf{A} is known as the
Bose-Mesner algebra of $Y$. Since the matrices $A_i$ commute, they
can be diagonalized simultaneously.

We now recall the dual Bose-Mesner algebra of $Y$. Given a base
vertex $\alpha\in V$, for all integers $i$ define
$E^*=E^*(\alpha)\in Mat_V(C)$ ($0\leq i\leq d $) to be the
diagonal matrix with $(\beta, \beta)$-entry
\begin{equation}\label{ee1}
\bigl(E_{i}^*)_{\beta, \beta}\;=\;\cases{1 & if $\;(\alpha,
\beta)\in R_i$,\cr 0 & otherwise\cr}\qquad \qquad (\alpha\in V).
\end{equation}
The matrix $E_i^*$ is called the $i$-th dual idempotent of $Y$
with respect to $\alpha$. We shall always set $E_i^*= 0$ for $i <
0$ or $i> d$. From the definition, the dual idempotents satisfy
the relations
\begin{equation}\label{e1}
\sum_{i=0}^{d}E_{i}^*=I,\;\;\ E_{i}^*E_{j}^*=\delta_{ij}E_{i}^*
\;\;\;\;\ 0\leq i,j\leq d.
\end{equation}
It follows that the matrices $E_{0}^*$, $E_{1}^*$, ..., $E_{d}^*$
form a basis for a subalgebra $\textsf{A}^* =
\textsf{A}^*(\alpha)$ of $Mat_V(c)$. $\textsf{A}^*$ is known as
the dual Bose-Mesner algebra of
$Y$ with respect to $\alpha$. \\
\textbf{Definition 4}\label{def2} (Terwilliger algebra) Let the
scheme $Y=(V,{\{R_i\}}_{0\leq i\leq d})$ be as in definition 1,
pick any $\upsilon\in V$, and let
$\textsf{T}=\textsf{T}(\upsilon)$ denote the subalgebra of
$Mat_V(C)$ generated by the Bose-Mesner algebra $\textsf{A}$ and
the dual Bose-Mesner algebra ${\textsf{A}}^*$. The algebra
$\textsf{T}$ is called Terwilliger algebra of $Y$ with respect to
$\upsilon$.

Let $W=C^V$ denote the vector space over $C$ consisting of column
vectors whose coordinates are indexed by $V$ and whose entries are
in $C$. We endow $W$ with the Hermitian inner product $\langle  ,
\rangle$ which satisfies $\langle u ,v \rangle=u^t\bar{v}$ for all
$u, v \in W$ , where $t$ denotes the transpose and - denotes the
complex conjugation. For all $\beta\in V$, let $\ket{\beta}$
denote the element of $W$ with a $1$ in the $\beta$ coordinate and
$0$ in all other coordinates. We observe $\{\ket{\beta} : \beta\in
V\}$ is an orthonormal basis for $W$. Using (\ref{ee1}) we have
\begin{equation}\label{Ter1}
W_i=E_i^*W = span\{\ket{\beta} : \beta\in V, (\alpha,\beta)\in
R_i\}, \;\;\;\ 0\leq i\leq d.
\end{equation}
Now using the relations (\ref{e1}), one can show that the operator
$E_i^*$ projects  $W$  onto $W_i$, thus we have
\begin{equation}\label{tel}
W=W_0\oplus W_1\oplus\cdots \oplus W_d.
\end{equation}

In \cite{js1}, CTQW on some special kinds of underlying graphs of
$P$-polynomial association schemes has been investigated. It is
shown in \cite{bannai} that in the case of $P$-polynomial
association schemes, $A_i=p_i(A) (0\leq i\leq d)$, where  $p_i$ is a
polynomial of degree $i$ with real coefficients. In particular, $A$
generates the Bose-Mesner algebra. Moreover, for a $P$-polynomial
scheme, there is a quantum decomposition for adjacency matrix of the
underlying graph, where in \cite{js1}, this property has been
employed for investigation of CTQW via spectral distribution
associated with adjacency matrix. In fact, for $P$-polynomial
schemes a
quantum decomposition for adjacency matrix can be defined by the following lemma\\
\textbf{Lemma  (Terwilliger \cite{T1})}.  Let $\Gamma$ denote an
underlying  graph of a $P$-polynomial association scheme with
diameter $d$. Fix any vertex $\alpha$ of $\Gamma$, and write
$E_i^* =E_i^*(\alpha)$ ($0\leq i\leq d$), $A_1=A$ and $\textit{T}
= \textit{T }(\alpha)$. Define $A^{-} = A^{-}(\alpha)$, $A^{0} =
A^{0}(\alpha)$, $A^{+} = A^{+}(\alpha)$ by
\begin{equation}\label{dec1}
A^{-}=\sum_{i=1}^{d}E_{i-1}^*AE_{i}^*, \;\;\;\;\
A^{0}=\sum_{i=1}^{d}E_{i}^*AE_{i}^*, \;\;\;\;\ A^{+}=
\sum_{i=1}^{d}E_{i+1}^*AE_{i}^*.
\end{equation}
Then
\begin{equation}
A=A^{+}+A^{-}+A^{0},
\end{equation}
 where, this is quantum decomposition of adjacency matrix $A$ such
 that,
\begin{equation}
   (A^{-})^{t} = A^{+}, \;\;\;\;\;\  (A^{0})^t =A^{0},
\end{equation}
which can be verified easily.

 Note that the above lemma is true only in the cases
of $P$-polynomial association schemes. In this paper we will
construct some underlying graphs of association schemes for which
the corresponding Bose-Mesner algebras are generated by $n$
commuting operators. Hereafter, we will refer to these types of
association schemes as $n$-variable $P$-polynomial association
schemes. As a generalization of the above lemma to $n$-variable
$P$-polynomial association schemes, one can define raising, lowering
and flat operators as in (\ref{dec1}) with respect to each generator
of Bose-Mesner algebra. In particular, for association scheme
derived from $Z_m\times Z_m$, the corresponding Bose-Mesner algebra
is generated by two commuting operators $A_z$ and $A_{\bar{z}}$
($A_{\bar{z}}=A^t_z$), i.e., $A_{kl}=p_{kl}(A_z,A_{\bar{z}})$, where
$p_{kl}$ is a polynomial of degree $k+l$ with real coefficients. The
raising, lowering and flat operators are defined as in (\ref{dec1})
with respect to each generator of Bose-Mesner algebra. Explicitly we
have
$$A_z^{+}:=\sum_i E_{i+1}^*A_zE_{i}^*  \;\;\;\;\;\;\
A_{\bar{z}}^{+}:=\sum_i E_{i+1}^* A_{\bar{z}}E_{i}^*,$$
$$ A_{\bar{z}}^{-}:=(A_z^{+})^t \;\;\;\;\;\;\;\;\;\;\;\;\ A_{z}^{-}:=(A_{\bar{z}}^{+})^t \;\;\;\ \mbox{and}$$
\begin{equation}\label{rais}
A_z^{0}:=\sum_i E_{i}^*A_zE_{i}^*  \;\;\;\;\;\;\
A_{\bar{z}}^{0}:=\sum_i E_{i}^* A_{\bar{z}}E_{i}^*.
\end{equation}
Similar to $P$-polynomial association schemes, we have
\begin{equation}
A_z=A_z^{+} +A_z^{-}+A_z^{0},\;\;\ A_{\bar{z}}=A_{\bar{z}}^{+}
+A_{\bar{z}}^{-}+A_{\bar{z}}^{0}.
\end{equation}
\section{construction of some translation invariant association
schemes } In this section, we construct two types of finite
underlying graphs of association schemes from finite abelian group
$Z_m\times Z_m$ ($m\geq3$) and two copies of finite hexagonal
lattices, such that in the limit of the large size of the graphs,
the underlying graphs tend to infinite graphs on root lattice $A_2$
and honeycomb one, respectively. We will show that the corresponding
Bose-Mesner algebras are generated by two commuting operators, in
particular all elements of Bose-Mesner algebras are two-variable
polynomials of the generators. We will refer to these schemes as
two-variable $P$-polynomial association schemes. To our purpose,
first we give some definitions.

\textbf{Definition 6} Let $A$ be a finite multiplicative abelian
group and $R=\{R_0,...,R_r\}$ a collection of $r+1$ distinct
relations on $A$ forming a partition of the cartesian power $A^2$.
If $(x,y)\in R_i$ implies $(ax,ay)\in R_i$ for all $a\in A$ and
$i=0,1,...,r$, then $P$ is called translation invariant.

\textbf{Definition 7} A partition $P=\{P_0,...,P_r\}$ of an
abelian group $A$ is called a blueprint \cite{Ass.sch.} if\\
(1) $P_0=\{e\}$ ($e$ is the identity of the group),\\
(2) for $i=1,...,r$, if $x\in P_i$ then $x^{-1}\in P_i$ (i.e.,
$P_i={P_i}^{-1}$),\\
(3) there are integers $q_{ij}^k$ such that if $y\in P_k$ then
there are precisely $q_{ij}^k$ elements $x\in P_i$ such that
$x^{-1}y\in P_j$.

 Now let $A$ be an abelian group, and $P=\{P_0,...,P_r\}$ be a
blueprint of $A$. Let $\Gamma(P)=\{R_0,...,R_r\}$ be the set of
relations
\begin{equation}\label{relation}
R_i=\{(x,y)\in A^2 \;\ | \;\  x^{-1}y\in P_i\},
\end{equation}
on $A$. One can notice that, if $P_i$ is a generating set for the
group $A$, then the underlying graph $\Gamma=(A,R_i)$ is called a
Cayley graph on $A$. From (\ref{relation}), it can be easily seen
that $R=\{R_0, ..., R_r\}$ forms a translation invariant partition
of $A^2$, where $R_0$ is diagonal relation. Also from condition
$(2)$ in definition $7$, $(x,y)\in R_i$ implies that $(y,x)\in
R_i$, i.e., $R_i^{-1}=R_i$.
 \subsection{construction of two-variable $P$-polynomial association schemes from $Z_m\times Z_m$ ($m\geq 3$)}
 First we choose the ordering of elements of $Z_m \times
Z_m$ as follows
\begin{equation}\label{ordering}
V=\{e,a,...,a^{m-1},b,ab,...,a^{m-1}b,...,b^{m-1},ab^{m-1},...,a^{m-1}b^{m-1}\},
\end{equation}
where $a^m=b^m=e$. We use the notation $(k,l)$ for the element
$a^kb^l$ of the group. Clearly, $(k,l)(k',l')=(k+k',l+l')$ and
$(k,l)^{-1}=(-k,-l)$. Then the vertex set $V$ of the graph will be
$\{(k,l) : k,l\in\{0,1,...,m-1\}\}$. Now we choose generating set
\begin{equation}\label{P1}
P_{10}=\{(1,0),(0,1),(m-1,m-1)\},
\end{equation}
 for $Z_m\times Z_m$. With this choice, we obtain Cayley graph $\Gamma=(V,
R_{10})$, where $V=Z_m \times Z_m$ and $R_{10}$ is defined by
(\ref{relation}).  Now, we obtain the orbits of Weyl group
 $S_3$ (all possible permutations of $(1,0)$, $(0,1)$ and $(m-1,m-1)$). Then,
 the orbits
\begin{equation}\label{orbit}
 P_{kl}:=O((k,-l)),
\end{equation}
form a partition $P$ for $Z_m\times Z_m$, where $P_{00}=\{(0,0)\}$
(in this case, $P$ is called homogeneous). Therefore, by using
(\ref{relation}), we obtain a coloring for the Cayley graph $(V,
R_{10})$ (with $R_{10}\neq R^{-1}_{10}$). Clearly, for the
relations $R_{kl}$ defined by (\ref{relation}) we have,  $\pi
R_{kl}{\pi}^{-1}=R_{kl}$ for every $\pi\in S_3$, i.e.,
$((x_1,x_2),(y_1,y_2))\in R_{kl}$ iff $(\pi (x_1,x_2),\pi
(y_1,y_2))\in R_{kl}$.

Moreover, since any product of two orbits $P_{k_1k_2}$ and
$P_{l_1l_2}$ is invariant under symmetric group $S_3$, the set of
orbits (consequently  the set of relations $R_{k_1k_2}$) is
closed under multiplication.  Also, if we use the notation
$i=(i_1,i_2), j=(j_1,j_2)$ and $k=(k_1,k_2)$, it can be easily
shown that, for $((x,x'),(y,y'))\in R_{k_1k_2}$, the intersection
number
$$p^k_{ij}=|\{(z,z'): ((x,x'),(z,z'))\in R_{i_1i_2},\;\ ((z,z'),(y,y'))\in
R_{j_1j_2} \}|$$
\begin{equation}
=|\{(z,z'): (z-x,z'-x')\in P_{i_1i_2},\;\ (y-z,y'-z')\in
P_{j_1j_2} \}|,
\end{equation}
is independent of the choice of $((x,x'),(y,y'))\in R_{k_1k_2}$.
Therefore, the relations $R_{kl}$ define an abelian association
scheme (not necessarily symmetric) on $Z_m\times Z_m$, where in the
regular representation of the group, for the corresponding adjacency
matrices  we have
\begin{equation}\label{adjacency}
A_{k,l}=\sum_{g\in P_{k,l}}g.
\end{equation}

From (\ref{orbit}) and (\ref{adjacency}), it follows that the
adjacency matrices satisfy the following recursion relations
$$A_{10}A_{k,l}=A_{k+1,l}+A_{k,l-1}+A_{k-1,l+1}\;\ ,$$
\begin{equation}\label{adj.recurs}
A_{01}A_{k,l}=A_{k-1,l}+A_{k,l+1}+A_{k+1,l-1}\;\ ,
\end{equation}
where, $A_{00}=I$, $A_{10}$ and $A_{01}=A^t_{10}$ are the first
adjacency matrices. In fact, the following two matrices
\begin{equation}\label{Az}
A_z:=S_1+S_2+(S_1S_2)^{-1}\;\ \mbox{and}\;\ A_{\bar{z}}:=(A_z)^t,
\end{equation}
generate the whole Bose-Mesner algebra of above constructed
association schemes. In particular,
$A_{kl}=p_{kl}(A_z,A_{\bar{z}})$, where $p_{kl}$ is a polynomial
of degree $k+l$ with real coefficients. We will refer to these
types of association schemes as two-variable $P$-polynomial
association schemes.

We illustrate the construction of underlying graph in simplest
case
$m=3$ in the following\\
\textbf{Example: case $m=3$}\\
 From (\ref{orbit}), the orbits of Weyl group $S_3$ are obtained as
$$P_{00}=\{(0,0)\},\;\ P_{10}=O((1,0))=\{(1,0),(0,1),(m-1,m-1)\},$$
\begin{equation}\label{P3}
P_{01}:=O((0,-1))=\{(2,0),(0,2),(1,1)\},\;\
P_{11}:=O((1,-1))=\{(1,2),(2,1)\}.
\end{equation}

Now  by using (\ref{relation}), one can obtain the relations
$R_{k_1k_2}$, for $(k_1,k_2)\in\{(0,0),(1,0),(0,1),(1,1)\}$. Also,
it can be verified that $\Gamma(P)=\{R_{k_1k_2}\}$ is an abelian
association scheme. The basis of Bose-Mesner algebra and dual
Bose-Mesner algebra are
$$A_{00}=I_9,\;\ A_{10}=S_1+S_2+(S_1S_2)^2,\;\ A_{01}=S^2_1+S^2_2+S_1S_2,\;\
A_{11}=S_1S^2_2+S^2_1S_2 \;\ \mbox{and}$$
$$E_{00}^*={E'_0}^*\otimes {E'_0}^*, \;\;\;\ E_{10}^*={E'_0}^*\otimes {E'_1}^*+{E'_1}^*\otimes
{E'_0}^*+{E'_2}^*\otimes {E'_2}^*,\;\;\;\ E^*_{01}={E'_0}^*\otimes
{E'_2}^*+{E'_2}^*\otimes {E'_0}^*+{E'_1}^*\otimes {E'_1}^*,$$
\begin{equation}\label{dualBM}
 {E_{11}}^*={E'_1}^*\otimes {E'_2}^*+{E'_2}^*\otimes {E'_1}^*,
\end{equation}
respectively, where
\begin{equation}
({E'_i}^*)_{yy}= \delta_{yi}, \;\;\;\ i=0,1,2.
\end{equation}

The adjacency matrices are written in terms of $A_z$ and
$A_{\bar{z}}$ as follows
\begin{equation}\label{m3}
A_{00}=I,\;\ A_{10}=A_z,\;\ A_{01}=A_{\bar{z}},\;\
A_{11}=\frac{1}{3}(A_zA_{\bar{z}}-3).
\end{equation}

One can notice that, the set $\{S_1,S_2\}$ is a generating set for
$Z_m\times Z_m$, i.e., the elements of the group in the regular
representation are of the form $(k,l)\doteq S_1^{k}S_2^{l}$, for
$k,l\in \{0,1,...,m-1\}$. If we represent $S_j$ as $S_j\doteq
e^{2\pi ix_j/m}$, $ x_j \in \{0,1,...,m-1\}$, then we have
$S_jS_k=e^{2\pi i(x_j+x_k)/m}$, so the multiplication in the
generating set $\{S_i ,i=1,2\}$ is equivalent to the addition in the
set $\{x_i ,i=1,2\}$. In the additive notation, $A_z$ is written as
\begin{equation}
 A_z=e^{2\pi ix_1/m}+e^{2\pi ix_2/m}+e^{-2\pi i(x_1+x_2)/m},
\end{equation}
 so, clearly $ \{(x_1,x_2,x_{3}=-(x_1+x_2))\;\ : \;\ x_i\in Z_m\}$ is a finite sequence of
 triples such that in the limit of large $m$ tends to the
root lattice $A_2$.
\subsubsection{finite hexagonal lattice}
The underlying graphs of two-variable $P$-polynomial association
schemes constructed in previous section are directed graphs since
the relation $R_{10}$ is non-symmetric. In this section, in order
to obtain undirected (symmetric) underlying graphs of two-variable
$P$-polynomial association schemes, we symmetrize the above
constructed graphs of previous section. To do so, we choose a
suitable union of the orbits such that the new partition $Q$ is
symmetric in the sense that $Q_{kl}=Q^{-1}_{kl}$, for all $(k,l)$.
In another words, we construct a blueprint from partition $P$, by
symmetrization. Such a symmetrization conserves the property of
being association scheme, because the union of the orbits is still
invariant under the action of symmetric group. In appendix $A$ of
\cite{js1}, such a symmetrization method is used for group
association schemes.

 Therefore, we construct the new underlying graph of association scheme,
by choosing the generating set $Q_{10}$ as follows
\begin{equation}
Q_{10}=P_{10}\cup
P_{01}=\{(1,0),(0,1),(1,1),(m-1,0),(0,m-1),(m-1,m-1)\}.
\end{equation}
 With this choice, the adjacency matrix of underlying graph is
\begin{equation}
A=A_z+A_{\bar{z}},
\end{equation}
where, $A_z$ and $A_{\bar{z}}$ are defined in (\ref{Az}). Clearly,
the new graph can be viewed as finite hexagonal lattice.
  In the following, we give the symmetric partition $Q$ and corresponding adjacency matrices of underlying
graph for $m=3$.\\
\textbf{Example: case $m=3$}\\
Using (\ref{P3}), the new partition $Q$ is given as
$$Q_{00}=\{(0,0)\},\;\;\ Q_{10}=\{(1,0),(0,1),(1,1),(m-1,0),(0,m-1),(m-1,m-1)\},$$
\begin{equation}
Q_{11}=P_{1,1}=\{(1,2),(2,1)\},
 \end{equation}
 and for the adjacency matrices, we
have
\begin{equation}
A_{00}=I,\;\;\
A_{10}=S_1+S_2+(S_1S_2)^2+(S_1)^2+(S_2)^2+S_1S_2,\;\;\
A_{11}=S_1S_2^2+S_1^2S_2.
 \end{equation}

Clearly the new constructed graphs are also underlying graphs of
two-variable $P$-polynomial association schemes. For example in the
case of $m=3$ we can write
\begin{equation}\label{m3}
A_{00}=1,\;\ A_{10}=A_z+A_{\bar{z}},\;\
A_{11}=\frac{1}{3}(A_zA_{\bar{z}}-3),
\end{equation}

In section $6$, we will investigate the behavior of CTQW on these
undirected graphs via spectral method, so we need to know the
spectrum of  adjacency matrix $A$.  The spectrum of $A_z$ in
(\ref{Az}) can be easily determined as
\begin{equation}\label{Eig}
z_{ij}=\omega^i+\omega^j+\omega^{-(i+j)},\;\;\ \omega=e^{2\pi i/m};
\;\;\ i,j\in\{0,1,...,m-1\}.
\end{equation}
Then, from (\ref{Eig}) and that the spectrum of $A_{\bar{z}}$ is
complex conjugate of the spectrum of $A_z$, one can calculate the
spectrum of $A$ as follows
  \begin{equation}\label{seig}
 \lambda_{kl}=z_{kl}+z^*_{kl}=2(\cos(2\pi k/m)+\cos(2\pi l/m)+\cos(2\pi(k+l)/m)).
 \end{equation}
  \subsection{construction of association scheme from two copies of hexagonal lattice}
We extend the group $Z_m \times Z_m$ by direct product with $Z_2$
and obtain $Z_2 \times Z_m \times Z_m$ as a vertex set for
underlying graph of association scheme that we want to construct. As
regards the argument of section $2$, we know that, finite honeycomb
lattice is equivalent to two copies of finite hexagonal lattice (see
Fig.2), therefore we define the adjacency matrix $A$ corresponding
to finite honeycomb lattice, such that $A^2$ gives us
$A_{hexagonal}$, the adjacency matrix of finite hexagonal lattice.
That is we have
\begin{equation}\label{honey}
A=\sigma_+\otimes B^t+\sigma_-\otimes B,
\end{equation}
where, $B=I+S_1+ S_2^{-1}$. Clearly
$B^tB=BB^t=S_1+S_2+S_1S_2+S_1^{-1}+S_2^{-1}+(S_1S_2)^{-1}=A_{hexagonal}$.

 By computing the powers of adjacency matrix $A$, one
can construct other adjacency matrices associated with an
association scheme (not necessarily $P$-polynomial).
Unfortunately, in this case we are not able to construct the
association scheme via a systematic procedure based on group
theoretical approach as in the case of finite hexagonal lattice
(this shows the preference of group theoretical approach). Also,
it should be noted that, in this case the association scheme is
defined in terms of matrices (see third definition of an
association scheme in \cite{Ass.sch.}). For example we give the
adjacency matrices of Bose-Mesner
algebra for $m=3$.\\
 \textbf{case $m=3$}\\
 The adjacency matrices of Bose-Mesner algebra are written as\\
$$A_0=I_2\otimes I_9 ,\;\ A_1=\sigma_+\otimes B^t+\sigma_-\otimes B ,\;\ A_2=I_2 \otimes A_{tri.},$$
$$A_3=\sigma_+\otimes
(S_1+S^2_2+S_1S_2+(S_1S_2)^2+S_1^2S_2+S_1S_2^2)+\sigma_-\otimes
(S_1^2+S_2+(S_1S_2)^2+S_1S_2+S_1^2S_2+S_1S_2^2),$$
\begin{equation}
A_4=I_2\otimes (S_1S_2^2+S_1^2S_2).
\end{equation}
One can see that $A_i$ for $i=1,...,4$ are symmetric and
$\sum_{i=0}^4 A_i=J_{18}$. Also it can be verified that,
$\{A_i,\;\ i=1,...,4\}$ is closed under multiplication and
therefore, the set of matrices $A_0,...,A_4$  form a symmetric
association scheme. We give only the following multiplications of
adjacency matrices, where we will use them later
\begin{equation}\label{rel3}
A_1^2=3A_0+A_2,\;\;\ A_1A_2=2A_1+2A_3,\;\;\ A_1A_3=2A_2+3A_4,\;\;\
A_1A_4=A_3.
\end{equation}
We will denote the graph constructed as above by $\Gamma_s$. It is
notable that, in the limit of large $m$, the graph $\Gamma_s$ can
be viewed as a graph with vertices belonging to honeycomb lattice.
In fact, starting from site $e$ of a hexagonal lattice, the
generators $cI$, $cS_1^{-1}$ and $cS_2$ ($c^2=1$), generate the
honeycomb lattice (See Fig.2). From Fig.2 one can see that, moving
on the honeycomb lattice by steps of length two, is equivalent to
moving on hexagonal lattice by steps of length one.

One can notice that, the graph $\Gamma_s$ is a bipartite graph and
has supersymmetric structure in the sense of
Ref.\cite{supersymetry}, where we discuss the supersymmetric
structure of $\Gamma_s$ in appendix.

\subsection{Stratification}
In this section, first we recall some of the main features of
stratification for underlying graphs of association schemes (see for
example \cite{js1}) and then stratify the underlying graphs of
association schemes constructed in previous subsections.

Let $V$ be the vertex set of an underlying graph $\Gamma$ of
association scheme. For a given vertex $\alpha\in V$, the set of
vertices having relation
 $R_i$ with $\alpha$ is denoted by
 $\Gamma_i(\alpha)=\{\beta\in V:
(\alpha, \beta)\in R_i\}$. Therefore, the vertex set $ V$ can be
written as disjoint union of $\Gamma_i(\alpha)$ for
$i=0,1,2,...,d$ (where, $d$ is diameter of the corresponding
association scheme), i.e.,
 \begin{equation}\label{asso1}
 V=\bigcup_{i=0}^{d}\Gamma_{i}(\alpha).
 \end{equation}
Now, we fix a point $o\in V$ as an origin of the underlying graph,
called reference vertex. Then, the relation (\ref{asso1})
stratifies the graph into a disjoint union of strata (associate
classes) $\Gamma_{i}(o)$.

 With each stratum $\Gamma_{i}(o)$ we associate a unit vector
$\ket{\phi_{i}}$ in $l^2(V)$ (called unit vector of $i$-th
stratum) defined by
\begin{equation}\label{unitv}
\ket{\phi_{i}}=\frac{1}{\sqrt{a_{i}}}\sum_{\alpha\in
\Gamma_{i}(o)}\ket{\alpha}\in E_i^*W ,
\end{equation}
where, $\ket{\alpha}$ denotes the eigenket of $\alpha$-th vertex at
the associate class $\Gamma_{i}(o)$ and $a_i=|\Gamma_{i}(o)|$. For
$0\leq i\leq d$ the unit vectors $\ket{\phi_{i}}$ of
Eq.(\ref{unitv}) form a basis for irreducible submodule of
corresponding Terwilliger algebra with maximal dimension denoted by
$W_0$ (\cite{T1}, Lemma 3.6). The closed subspace of $l^2(V)$
spanned by $\{\ket{\phi_{i}}\}$ is denoted by $\Lambda(G)$. Since
$\{\ket{\phi_{i}}\}$ becomes a complete orthonormal basis of
$\Lambda(G)$, we often write
\begin{equation}
\Lambda(G)=\sum_{i}\oplus \textbf{C}\ket{\phi_{i}}.
\end{equation}

In the graphs constructed from $Z_m\times Z_m$, the vertex set is
$V=\{(k,l) : k,l\in\{0,1,...,m-1\}\}$. Therefore, for a given vertex
$(m,n)\in V$, $\Gamma_{kl}((m,n))=\{(m',n'): ((m,n),(m',n'))\in
R_{kl}\}$ is equivalent to
\begin{equation}
\Gamma_{kl}((m,n))=\{(m',n'): (m'-m,n'-n)\in O((k,-l))\},
\end{equation}
where, $O((k,-l))$ denote the orbits of Weyl group corresponding to
the finite lattice. Now, we fix the vertex $(0,0)\in V$ as an origin
of the underlying graph, called reference vertex. Then, the relation
(\ref{asso1}) stratifies the graph into a disjoint union of
associate classes $\Gamma_{kl}((0,0))$. Then, the unit vectors
(\ref{unitv}) are written as
\begin{equation}\label{unitv1}
\ket{\phi_{kl}}=\frac{1}{\sqrt{a_{kl}}}\sum_{(m,n)\in
\Gamma_{kl}((0,0))}\ket{m,n},
\end{equation}
where, $a_{kl}=|\Gamma_{kl}((0,0))|$. In section $6$, we will deal
with the CTQW on the constructed underlying graphs, where the strata
$\{\ket{\phi_{kl}}\}$ span a closed subspace (irreducible submodule
of corresponding Terwilliger algebra with maximal dimension called
walk space), where the quantum walk remains on it forever.

For reference state $\ket{\phi_{00}}=\ket{00}$
 we have
\begin{equation}\label{Foc1}
A_{kl}\ket{\phi_{00}}=\sum_{(m,n)\in \Gamma_{kl}((0,0))}\ket{m,n}.
\end{equation}
 Then by using unit vectors (\ref{unitv1}) and (\ref{Foc1}) one can
 see that
\begin{equation}\label{Foc2}
A_{kl}\ket{\phi_{00}}=\sqrt{a_{kl}}\ket{\phi_{kl}}.
\end{equation}

In the case of finite honeycomb lattice, we have two sets of odd
and even vertices, i.e., $V=V_o+V_e$, where $V_o$ is the set of
odd vertices defined by $V_o=\{(1;k,l): k,l\in\{0,1,...,m-1\}\}$
and $V_e$ is the set of even vertices defined by $V_e=\{(0;k,l):
k,l\in\{0,1,...,m-1\}\}$. We define stratum $\Gamma_i(u;k,l)$ as
\begin{equation}
\Gamma_i(u;k,l)=\{(v,k',l'): (A_i)_{((u;k,l),(v;k',l'))=1}\},
\end{equation}
where, $u,v\in{0,1}$ and $k,l,k',l'\in\{0,1,...,m-1\}$. Now, we
fix the vertex $(0;0,0)\in V$ as an origin of the underlying
graph. Then, the relation (\ref{asso1}) stratifies the graph into
a disjoint union of associate classes $\Gamma_{i}((0;0,0))$ and
the relations (\ref{unitv1}), (\ref{Foc1}) and (\ref{Foc2}) are
satisfied by replacing $\Gamma_i(0,0)$ with $\Gamma_{i}((0;0,0))$.

One should notice that, these types of stratifications are different
from the one based on distance, i.e., it is possible that two strata
with the same distance from starting site posses different
probability amplitudes.
\section{Spectral distribution}
In this section we give a brief review of spectral distributions for
operators. Although the  spectrum of underlying graphs on which we
study CTQW, is easily evaluated and so CTQW can be investigated
without spectral distribution approach, but in the limit of large
size of the finite graphs, the best approach for calculating
expected values of adjacency matrices  is spectral distribution one.
As we will see later, based on spectral distribution, one can
approximate the behavior of the CTQW on infinite graphs with finite
ones via stationary phase approximation method. Also, the spectral
distribution approach is  the best method for studying central limit
theorems for quantum walks on graphs, see for example \cite{konno},
\cite{GJS}.

In \cite{js1} and \cite{jss}, CTQW on underlying graphs of QD type
is investigated via spectral distribution, where the spectral
measures associated with the adjacency matrices are single variable.
In the case of $n$-variable $P$-polynomial association schemes,
spectral measures are $n$-variable functions. Therefore, in the
following we generalize the discussions in \cite{js1} and \cite{jss}
to the case of $n$-variable $P$-polynomial association schemes.

 It is well known that, for every set of commuting operators
$(A_{z_1},...,A_{z_n})$ and a reference state $\ket{\phi_0}$, it can
be assigned a distribution measure $\mu$ as follows
\begin{equation}\label{spp1}
\mu(z_1,...,z_n)=\braket{ \phi_0}{E(z_1,...,z_n)|\phi_0},
\end{equation}
 where
$E(z_1,...,z_n)=\sum_i|u_i^{(z_1,...,z_n)}\rangle\langle
u_i^{(z_1,...,z_n)}|$ is the operator of projection onto the common
eigenspace of $A_{z_1}$,...,$A_{z_n}$ corresponding to eigenvalues
$z_1$,...,$z_n$, respectively. Then, for any $n$-variable polynomial
$P(A_{z_1},...,A_{z_n})$ we have
\begin{equation}\label{spp2}
P(A_{z_1},...,A_{z_n})=\int...\int
P(z_1,...,z_n)E(z_1,...,z_n)dz_1...dz_n,
\end{equation}
where for discrete spectrum the above integrals are replaced by
summation. Using the relations (\ref{spp1}) and (\ref{spp2}), we
have
\begin{equation}\label{vv2}
\braket{\phi_{0}}{P(A_{z_1},...,A_{z_n})|\phi_0}=\int...\int
P(z_1,...,z_n) \mu(z_1,...,z_n)dz_1...dz_n.
\end{equation}
The existence of a spectral distribution satisfying (\ref{vv2}) is
a consequence of Hamburger's theorem, see e.g., Shohat and
Tamarkin [\cite{st}, Theorem 1.2].

 Actually the spectral analysis of operators  is an important issue in quantum mechanics, operator
theory and mathematical physics \cite{simon, Hislop}. As an
example $\mu(dx)=|\psi(x)|^2dx$
($\mu(dp)=|\widetilde{\psi}(p)|^2dp$) is a spectral distribution
which is  assigned to  the position (momentum) operator
$\hat{X}(\hat{P})$. Moreover, in general quasi-distributions are
the assigned spectral distributions of two hermitian non-commuting
operators with a prescribed ordering.
 For
example the Wigner distribution in phase space is the assigned
spectral distribution for two non-commuting operators $\hat{X}$
(shift operator) and $\hat{P}$ (momentum operator) with
Wyle-ordering among them \cite{Kim,Hai}.

\subsection{construction of orthogonal polynomials}
As regards the arguments of section $4$, the Bose-Mesner algebra
corresponding to two-variable $P$-polynomial association scheme
derived from orbits of Wyel group corresponding to finite hexagonal
lattice, is generated by $A_z$ and $A_{\bar{z}}$ defined by
(\ref{Az}). We assign the variables $z$ and $\bar{z}$ to $A_z$ and
$A_{\bar{z}}$, respectively. Then, in the limit of the large size of
the underlying graph, the recursion relations (\ref{adj.recurs})
define a set of two-variable polynomials $p_{k,l}$ with the first
polynomials and recursion relations as follows
$$ P_{0,0}=1,\;\;\ P_{1,0}=z,\;\;\ P_{0,1}=\bar{z}\;\ ,$$
$$zP_{k,l}=P_{k+1,l}+P_{k,l-1}+P_{k-1,l+1}\;\ ,$$
\begin{equation}\label{recursion}
\bar{z}P_{k,l}=P_{k,l+1}+P_{k-1,l}+P_{k+1,l-1}\;\ ,
 \end{equation}
where, the polynomials $P_{m,n}$ in (\ref{recursion}) are orthogonal
with respect to the constant measure $\mu(x_1,x_2)=1$ (where,
$z=e^{ix_1}+e^{ix_2}+e^{-i(x_1+x_2)}$), i.e., we have
\begin{equation}
\int_{0}^{2\pi}\int_{0}^{2\pi}
P_{m,n}P_{m',n'}dx_1dx_2=\delta_{m,m'}\delta_{n,n'}.
\end{equation}

From (\ref{adj.recurs}) and (\ref{Foc2}), it can be seen that, there
is a canonical isomorphism from the interacting Fock space of CTQW
(irreducible submodule of Terwilliger algebra with highest
dimension) on the symmetric underlying graphs of two-variable
$P$-polynomial association schemes derived from $Z_m\times Z_m$
(finite hexagonal lattice) onto the closed linear span of the
orthogonal polynomials generated by recursion relations
(\ref{recursion}). In fact, the adjacency matrices of non-symmetric
association schemes constructed from $Z_m\times Z_m$ in section $4$,
are equal to polynomials $P_{m,n}(A_z,A_{\bar{z}})$ and the
symmetrization of the association schemes is equivalent to
realification of two-variable polynomials $P_{m,n}(z,\bar{z})$.
Therefore, the adjacency matrices of symmetric association schemes
derived from $Z_m\times Z_m$, are of the form $P_{m,n}(z,\bar{z})$
if $P_{m,n}(z,\bar{z})$ is real or of the form
$P_{m,n}(z,\bar{z})+\bar{P}_{m,n}(z,\bar{z})$ if
$P_{m,n}(z,\bar{z})$ is complex.

It should be noted that, in the case of finite hexagonal lattice,
the polynomials $P_{k,l}$ are not independent. Also, it can be
shown that, these polynomials can be derived by using the raising
operators $A^{+}_z$ and $A^{+}_{\bar{z}}$ defined by (\ref{rais})
corresponding to symmetric underlying graphs. In the following, we
list the strata and corresponding polynomials in the order of
their first appearances as
$$|\phi_0\rangle \longrightarrow P_{0,0}$$
$$\ket{\phi_{1,0}}=A^{+}_z|\phi_0\rangle \rightarrow P_{1,0},\;\;\ \ket{\phi_{0,1}}= A^{+}_{\bar{z}}|\phi_0\rangle \rightarrow
P_{0,1},$$
$$\ket{\phi_{2,0}}=(A^{+}_z)^2|\phi_0\rangle \rightarrow
P_{2,0},\;\ \ket{\phi_{1,1}}= A^{+}_zA^{+}_{\bar{z}}|\phi_0\rangle
\rightarrow P_{1,1},\;\
\ket{\phi_{0,2}}=(A^{+}_{\bar{z}})^2|\phi_0\rangle \rightarrow
P_{0,2},$$
$$\ket{\phi_{3,0}}=(A^{+}_z)^3|\phi_0\rangle \rightarrow P_{3,0},
\ket{\phi_{2,1}}=(A^{+}_z)^2A^{+}_{\bar{z}}|\phi_0\rangle
\rightarrow
P_{2,1},\ket{\phi_{1,2}}=(A^{+}_{\bar{z}})^2A^{+}_z|\phi_0\rangle
\rightarrow P_{1,2},
\ket{\phi_{0,3}}=(A^{+}_{\bar{z}})^3|\phi_0\rangle \rightarrow
P_{0,3},$$
\begin{equation}\label{strpoly}
 ...
\end{equation}

  For the sake of clarity, we
construct
the polynomials in the simplest case $m=3$.\\
\textbf{Example: case $m=3$}\\
 In this case, we have $A^{+}_z=\sum_{i=0}^2 {E''}^*_{i+1}A_z{E''}^*_i$
 and $A^{+}_{\bar{z}}=\sum_{i=0}^2 {E''}^*_{i+1}A_{\bar{z}}{E''}^*_i$,
 where $A_z$ and $A_{\bar{z}}$ are given by (\ref{Az}) and the
 basis of dual Bose-Mesner algebra is given by
\begin{equation}
{E''}_0^*=E^*_0,\;\ {E''}_1^*=E^*_1+E^*_2,\;\ {E''}^*_2=E^*_3,
\end{equation}
where, $E^*_i$ for $i=0,1,2,3$ are given in (\ref{dualBM}). Now,
by using $A^{+}_z$ and $A^{+}_{\bar{z}}$, we obtain the following
states
$$\ket{\phi_{1,0}}=A^{+}_z\ket{00}=\ket{02}+\ket{20}+\ket{11}=A_z\ket{00},$$
$$\ket{\phi_{0,1}}=A^{+}_{\bar{z}}\ket{00}=\ket{01}+\ket{10}+\ket{22}=A_{\bar{z}}\ket{00},$$
\begin{equation}
\ket{\phi_{1,1}}=A^{+}_zA^{+}_{\bar{z}}\ket{00}=3(\ket{12}+\ket{21})=(A_zA_{\bar{z}}-3I)\ket{00}.
\end{equation}
Therefore, the two-variable orthogonal polynomials associated with
$\ket{\phi_{1,0}}$, $\ket{\phi_{0,1}}$ and $\ket{\phi_{1,1}}$ are
\begin{equation}
P_{1,0}=z, \;\ P_{0,1}=\bar{z}\;\ \mbox{and}\;\
P_{1,1}=z\bar{z}-3,
\end{equation}
respectively. Moreover, the polynomials $P_{m,n}$ are special
cases of orthogonal polynomials known as \textit{generalized}
Gegenbauer polynomials \cite{Per1}, \cite{Per2}. These
polynomials also can be derived from solving the schrodinger
equation for special case of completely integrable quantum
Calogero-Sutherland model of $A_n$ type  with constant potential,
which describes the mutual interaction of $N = n + 1$ particles
moving on the circle. The coordinates of these particles are
$x_j$,  $j=1,...,N$  and the Schrodinger equation reads as
\begin{equation}
H\Psi= E\Psi,\;\ H = -\frac{1}{2}\Delta , \;\;\;\
\Delta=\sum_{j=1}^N \frac{\partial^2}{\partial {x_j}^2}.
\end{equation}
The ground-state energy and (non-normalized) wavefunction are
\begin{equation}
E_0=0,\;\;\ \Psi_0(x_i)=1.
 \end{equation}
  The excited states depend on an
$n$-tuple of quantum numbers $m=(m_1,m_2,...,m_n)$:
\begin{equation}
H \Psi_m(x_i) = E_m \Psi_m,,\;\;\ E_m= 2(\lambda , \lambda),
\end{equation}
 where $\lambda$ is the highest weight
of the representation of $A_n$ labeled by $m$, i.e., $\lambda =
\sum_{i=1}^n m_ie_i$ and $e_i$ are the fundamental weights of
$A_n$. In fact, the eigenfunctions $\Psi_m$ are solutions to the
Laplace equation
\begin{equation}
-\Delta \Psi_m=E_m \Psi_m.
\end{equation}

Let us restrict ourselves to the case $A_2$. If we change the
variables as
\begin{equation}
z_1=e^{2ix_1}+e^{2ix_2}+e^{2ix_3},\;\
z_2=e^{2i(x_1+x_2)}+e^{2i(x_2+x_3)}+e^{2i(x_3+x_1)},\;\
z_3=e^{2i(x_1+x_2+x_3)},
\end{equation}
 then, in the center-of-mass frame
$(\sum_i x_i=0)$, the wavefunctions depend only on two variables
chosen as $z=z_1$ and $\bar{z}=z_2$ (in this case, $z_3=1$). With
this change of variables and using normalization for $\Psi_m$ such
that the coefficient at the highest monomial is equal to one, we
obtain the orthogonal polynomials $P_{m_1,m_2}$ with respect to the
constant measure $\Psi_0$ (the polynomials are correspond to exited
states) which satisfy the recursion relations (\ref{recursion}).
\section{CTQW on underlying graphs of two-variable $P$-polynomial association schemes via spectral method}
CTQW was introduced by Farhi and Gutmann in Ref.\cite{fg}. Let
$l^2(V)$ denote the Hilbert space of $C$-valued square-summable
functions on V (i.e., $\sum_i |\textit{f}_i|^2 <\infty$). With each
$\alpha\in V$ we associate a ket $\ket{\alpha}$, then
$\{\ket{\alpha},\;\ \alpha\in V \}$ becomes a complete orthonormal
basis of $l^2(V)$.

Let $\ket{\phi(t)}$ be a time-dependent amplitude of the quantum
process on  graph $\Gamma$. The wave evolution of the quantum walk
is
\begin{equation}
    i\hbar\frac{d}{dt}\ket{\phi(t)} = H\ket{\phi(t)},
\end{equation}
where assuming $\hbar = 1$,  and $\ket{\phi_{0}}$ be the initial
amplitude wave function of the particle, the solution is given by
$\ket{\phi_{0}(t)} = e^{-iHt} \ket{\phi_{0}}$. It is more natural to
deal with the Laplacian  of the graph defined by $L=A-D$ as
hamiltonian, where $D$ is a diagonal matrix with entries
$D_{jj}=\mbox{deg}(\alpha_j)$ (recall that $\mbox{deg}(\alpha_j)$ is
degree of the vertex $\alpha_j$ defined by the number of edges
incident to the vertex $\alpha_j$). This is because we can view $L$
as the generator matrix that describes an exponential distribution
of waiting times at each vertex. But on $d$-regular graphs, $D =
\frac{1}{d}I$, and since $A$ and $D$ commute, we get
\begin{equation} \label{eqn:phase-factor}
e^{-itH} = e^{-it(A-\frac{1}{d}I)} = e^{-it/d}e^{-itA},
\end{equation}
this introduces an irrelevant phase factor in the wave evolution.
In this paper we consider $L=A=A_1$. Therefore, we have
\begin{equation}
\ket{\phi_{0}(t)}=e^{-iAt}\ket{\phi_0}.
\end{equation}
 One approach for investigation of
CTQW on graphs is using the spectral distribution method. CTQW on
underlying graphs of $P$-polynomial association schemes has been
discussed exhaustively in \cite{js} via spectral method. In the
following we investigate CTQW on underlying graphs of two-variable
$P$-polynomial association schemes constructed in section $4$ using
spectral distribution method.
\subsection{CTQW on underlying graphs of two-variable $P$-polynomial
association schemes derived from $Z_m\times Z_m$}

In the graphs constructed from $Z_m\times Z_m$, the adjacency matrix
is written as $A_z+A_{\bar{z}}$ and so we assign polynomial
$z+\bar{z}$ to adjacency matrix. Then, by using the relation
(\ref{vv2}), the expectation value of powers of adjacency matrix $A$
over starting site $\ket{\phi_{00}}$ can be written as
\begin{equation}\label{vv3}
\braket{\phi_{00}}{A^m|\phi_{00}}=\int\int(z+\bar{z})^m
\mu(z,\bar{z})dzd\bar{z}, \;\;\;\;\ m=0,1,2,....
\end{equation}
In the case of underlying graphs of two-variable $P$-polynomial
association schemes, the adjacency matrices are two-variable
polynomial functions of $A_z$ and $A_{\bar{z}}$, hence using
(\ref{Foc2}) and (\ref{vv3}), the matrix elements $\label{cw1}
\braket{\phi_{kl}}{A^m\mid \phi_{00}}$ can be written as
$$
\braket{\phi_{kl}}{A^m\mid
\phi_{00}}=\frac{1}{\sqrt{a_{kl}}}\braket{\phi_{00}}{A_{kl} A^m\mid
\phi_{00}}
=\frac{1}{\sqrt{a_{kl}}}\braket{\phi_{00}}{P_{kl}(A_z,A_{\bar{z}})
A^m\mid \phi_{00}}$$
\begin{equation}\label{cww1}
=\frac{1}{\sqrt{a_{kl}}}\int_{R}\int_{R}(z+\bar{z})^{m}P_{kl}(z,\bar{z})\mu(z,\bar{z})dzd\bar{z},
\;\;\;\;\ m=0,1,2,... .
\end{equation}

 One of our goals in this paper is the evaluation of amplitudes
for CTQW on underlying graphs of two-variable $P$-polynomial
association schemes constructed in section $4$ via spectral
distribution method. By using (\ref{cww1}) we have
\begin{equation} \label{v4}
P_{kl}(t)=\braket{\phi_{kl}}{e^{-iAt}|\phi_{00}}=\braket{\phi_{kl}}{\phi_{00}(t)}=\frac{1}{\sqrt{a_{kl}}}\int_{R}\int_{R}e^{-i(z+\bar{z})t}P_{kl}(z,\bar{z})\mu(z,\bar{z})dzd\bar{z},
\end{equation}
where $\braket{\phi_{kl}}{\phi_{00}(t)}$ is the amplitude of
observing the particle at level $kl$ (stratum $\Gamma_{kl}((0,0))$)
at time $t$. One should notice that, as illustrated in section $5$,
the polynomials $p_{kl}(z,\bar{z})$ are obtained from realification
of \textit{generalized} Gegenbauer polynomials $P_{m,n}$ defined by
(\ref{recursion}). The conservation of probability $\sum_{k=0}{\mid
\braket{\phi_{kl}}{\phi_{00}(t)}\mid}^2=1$ follows immediately from
(\ref{v4}) by using the completeness relation of orthogonal
polynomials $P_{mn}(z,\bar{z})$. Obviously evaluation of
$\braket{\phi_{kl}}{\phi_{00}(t)}$ leads to the determination of the
amplitudes at sites belonging to the stratum $\Gamma_{kl}((0,0))$.

  Spectral distribution $\mu$ associated with the generators is defined as
 \begin{equation}\label{spec}
 \mu(z,\bar{z}) = \frac{1}{m^2} \sum_{k,l}
  \delta(z-z_{k,l})\delta(\bar{z}-z^*_{k,l}),
  \end{equation}
where $k,l \in \{0,1,...,m-1\}.$ Now using (\ref{v4}) and spectral
distribution (\ref{spec}), the probability amplitude of observing
the walk at stratum $\Gamma_{ij}((0,0))$ at time $t$ can be
calculated as
\begin{equation}\label{pk}
P_{ij}(t)=\frac{1}{m^2}\sum_{k,l}e^{-2it(\cos{2\pi {k}/m}+\cos{2\pi
{l}/m}+\cos{2\pi(k+l)/m})}p_{ij}(z_{k,l},z^*_{k,l}).
\end{equation}
In particular, the probability amplitude of observing the walk at
starting site at time $t$ is given by
\begin{equation}\label{p0}
P_{00}(t):=\frac{1}{m^2}\sum_{k,l}e^{-2it(\cos{2\pi {k}/m}+\cos{2\pi
{l}/m}+\cos{2\pi(k+l)/m})}.
\end{equation}

 \textbf{Example: case $m=3$.}\\
By using (\ref{Eig}), we obtain $z_{kl}\in\{0,3,
3\omega,3\omega^2\}$. Then by (\ref{spec}), spectral distribution is
calculated as
\begin{equation}
\mu(z,\bar{z})=\frac{1}{9} \{\delta(z-3)\delta(\bar{z}-3)+6
\delta(z)\delta(\bar{z})+\delta(z-3\omega)\delta(\bar{z}-3\omega^2)+
\delta(z-3\omega^2)\delta(\bar{z}-3\omega)\}.
\end{equation}
Therefore, by using (\ref{m3}) and (\ref{v4}), probability
amplitudes of observing the walk at starting site, stratum
$\Gamma_{10}((0,0))$ and $\Gamma_{11}((0,0))$ are calculated as
 $$ P_{00}(t)=\frac{1}{9}\{e^{-6it}+e^{3it}+6\},$$
$$P_{10}(t)=\frac{2}{3}\{e^{-6it}-e^{3it}\},$$
\begin{equation}
P_{11}(t)=\frac{2}{9} \{e^{-6it}+2e^{3it}-3\},
\end{equation}
respectively.

 At the limit of the large $m$, we obtain the root
lattice $A_2$ (hexagonal lattice). In the following we investigate
the CTQW on root lattice $A_2$ using spectral method.

\subsection{CTQW on hexagonal lattice}
In this subsection we give continuous measure in the limit of the
large size of the underlying graphs of symmetric two-variable
$P$-polynomial association schemes derived by $A_z+A_{\bar{z}}$ in
section $4$.

In the limit of large $m$, the roots
$z_{kl}=\omega^k+\omega^l+\omega^{-(k+l)}$  reduce to
$z_{kl}=e^{ix_1}+e^{ix_2}+e^{-i(x_1+x_2)}$ with
$x_1=\lim_{{k,m}\rightarrow \infty}2\pi k/m$ and
$x_2=\lim_{{l,m}\rightarrow \infty}2\pi l/m$ and the spectral
distribution given in (\ref{spec}), reduces to continuous
constant measure $\mu(x_1,x_2)=1/4\pi^2$.

Also, the measure $\mu$ can be given in terms of complex variables
$z$ and $\bar{z}$ as
$$\mu(z,\bar{z})=\int_0^{2\pi}\int_0^{2\pi}
dx_1dx_2\delta(z-(e^{ix_1}+e^{ix_2}+e^{-i(x_1+x_2)}))\delta(\bar{z}-(e^{-ix_1}+e^{-ix_2}+e^{i(x_1+x_2)}))=$$
\begin{equation}
=\frac{1}{4\pi^2\sqrt{-z^2\bar{z}^2+4(z^3+\bar{z}^3)-18z\bar{z}+27}}.
\end{equation}

Then, the probability amplitudes $P_{kl}(t)$ are given by
\begin{equation}\label{int1}
P_{kl}(t)=\langle\phi_{kl}|e^{-iAt}|\phi_{00}\rangle=
\int_{0}^{2\pi}\int_{0}^{2\pi}dx_1dx_2e^{-2it(\cos x_1+\cos x_2+\cos
(x_1+x_2))}p_{kl}(x_1,x_2),
\end{equation}
where, $A_{kl}=p_{kl}(x_1,x_2)$. In particular, the probability
amplitude of observing the walk at starting site at time $t$, is
calculated as
\begin{equation}\label{conp0}
P_{00}(t)=\int_{0}^{2\pi}\int_{0}^{2\pi}dx_1dx_2e^{-2it(\cos
x_1+\cos x_2+\cos (x_1+x_2))}.
\end{equation}
\subsubsection{asymptotic behavior of quantum walk on hexagonal lattice}
As regards argument of the end of section $4.1$, we can not obtain
an analytic expression for the amplitudes of the walk in the
infinite case, i.e., the integral appearing in the (\ref{int1}) is
difficult to evaluate, but we can approximate it for large time $t$
by using the stationary phase method. Studying the large time
behavior of quantum walk naturally leads us to consider the behavior
of integrals of the form
\begin{equation}\label{int4}
I(t)=\int_{-{\infty}}^{\infty} \int_{-{\infty}}^{\infty}dx_1dx_2
\textit{g}(\vec{x})e^{-it f(\vec{x})}
 \label{int}
\end{equation}
as $t$ tends to infinity. There is a well-developed theory of the
asymptotic expansion of integrals which allows us to determine,
very precisely, the leading terms in the expansion of the integral
in terms of simple functions of $t$ (such as inverse powers of
$t$). Our basic technique will be to evaluate this integral in
some approximation. The approximation we shall use will be the
semiclassical expansion, which amounts to the well-known
stationary phase approximation as applied to the path integral. In
this approximation, one can evaluate $I(t)$ in (\ref{int4})
asymptotically as follows
\begin{equation}\label{asympt}
\int\int dx_1 dx_2\textit{g}(\vec{x})e^{-itf(\vec{x})}\simeq
\sum_{\vec{a}}
\textit{g}(\vec{a})e^{-itf(\vec{a})}\frac{2\pi}{it}(DetA)^{-1/2},
\end{equation}
where, summation is over all stationary points $\vec{a}$ of function
$f(\vec{x})$ and $A$ is Hessian matrix corresponding to
$f(\vec{x})$. For more details about this approximation method, the
reader is refered to \cite{stat.book}.

Now, by using (\ref{asympt}) we can discuss the asymptotic
behavior of the amplitude of observing the quantum walk at
starting site at large time $t$, i.e., we deal with the integral
(\ref{int}),
 with $\textit{g}(\vec{x})=1$ and
\begin{equation}
f(\vec{x})=2(\cos x_1+\cos x_2+\cos(x_1+x_2)).
 \label{Eq:1}
\end{equation}
Therefore, for the asymptotic form of the amplitude $P_{00}(t)$ we
get
\begin{equation}
I_0(t)\simeq
\frac{\pi}{t}(\frac{1}{2\sqrt{3}}e^{-6it+i\pi/2}+\frac{1}{2}e^{2it}+\frac{1}{\sqrt{3}}e^{3it-i\pi/2}).
\label{Eq:asymp.}
\end{equation}
 The asymptotic behavior of probability amplitudes $P_{kl}(t)$ ($kl\neq 00$) at
large time $t$, can be evaluated similarly.

In order to obtain the asymptotic behavior of quantum walk on finite
hexagonal lattice at large time $t$, we can compare the finite
amplitude $P_{00}(t)$ (Eq.(\ref{p0})) and the continuous probability
(\ref{conp0}) at large time $t$. Therefore,  we calculate
numerically the difference of amplitudes of the walk on root lattice
$A_2$ with ones on finite hexagonal lattice, for large time $t$
\begin{equation}
\pi(m,t)=|I_0(t)-\frac{1}{m^2}\sum_{k,l=0}^{m-1}e^{-2it(\cos(\frac{2k\pi}{m})+\cos(\frac{2l\pi}{m})+\cos(\frac{2(k+l)\pi}{m}))}|.
\end{equation}

The result has been depicted in Fig.3. The figure shows that, the
difference $\pi(m,t)$ is limited to zero for $m$ larger than $ \sim
50$ and $t\sim 1000 $. Therefore, to study the behavior of
asymptotic quantum walk on finite hexagonal lattice, we can use
arithmetic, approximate it with root lattice $A_2$, and by using the
stationary phase method, study the behavior of asymptotic quantum
walk.
\subsection{CTQW on finite honeycomb lattice via spectral
method} In the graph $\Gamma_s$ constructed in section $4$ from two
copies of finite hexagonal lattices, we have
\begin{equation}
A=\sigma_+\otimes B^t+\sigma_-\otimes B,
 \end{equation}
where, $B=I+S_1^{-1}+S_2$. Therefore, the spectrum of $B$ is
calculated as
\begin{equation}\label{honeig}
z_{kl}=1+{\omega}^{-k}+{\omega}^l, \;\;\
k,l\in\{0,1,...,m-1\};\;\;\ \omega=e^{\frac{2\pi i}{m}},
\end{equation}
 and the eigenvalues of $A$ are given by
 \begin{equation}\label{spechony}
 \lambda_{kl}=\pm
|z_{kl}|=\pm\sqrt{3+2(\cos(2\pi k/m)+\cos(2\pi
l/m)+\cos(2\pi(k+l)/m))}.
\end{equation}
Now we apply spectral method by assigning variables $z_1$ and
$z_2$ to $B$ and $B^t$, respectively and a new variable $z$ for
$\sigma_+$ and $\sigma_-$ commonly. Then, the variable assigned to
adjacency matrix $A$ will be $|z_1|z+|z_2|(z-1)$. Clearly we have
$z_1 = z_2^*$ and so $|z_1|=|z_2|$. Therefore, we will have
\begin{equation}
A=|z_1|(2z-1).
\end{equation}
 The spectral distribution associated with adjacency matrix $A$ is
 given by
\begin{equation}\label{measurehony}
\mu(z_1,\bar{z_1};z)=\frac{1}{2}\mu(z_1,\bar{z_1})(\delta(z)+\delta(z-1)),
\end{equation}
where,
\begin{equation}
\mu(z_1,\bar{z_1})=\frac{1}{m^2}\sum_{k,l}
\delta(z_1-z_{kl})\delta(\bar{z_1}-z^*_{kl}).
\end{equation}
Then, the probability amplitude of observing the walk at stratum
$\Gamma_k(0;00)$ at time $t$ is calculated as follows
\begin{equation}\label{amph}
 P_k(t)=\int\int\int e^{-i(|z_1|(2z-1))t}p_k(z_1;z)\mu(z_1,\bar{z_1};z)d{z_1}d{\bar{z_1}}dz ;\;\;\
 k=1,2,3,4,
 \end{equation}
 where, $A_k=p_k(z_1;z)$. In particular, the probability amplitude of observing the walk at
starting site at time $t$ is calculated as
\begin{equation}\label{pfh}
P_0(t)=\int\int\int
e^{-i(|z_1|(2z-1))t}\mu(z_1,\bar{{z_1}};z)d{z_1}d\bar{{z_1}}dz
=\frac{1}{m^2}\sum_{k,l=0}^{m-1} \cos(\sqrt{3+2(cos 2\pi k/m+cos
2\pi l/m)})t.
\end{equation}

For the sake of clarity, in the following we give details for the case $m=3$.  \\
\textbf{Example: case $m=3$}\\
From the relations (\ref{rel3}), we have\\
$$A_0=1,\;\;\ A_1=p_1(z_1;z)=|z_1|(2z-1),\;\;\
A_2=p_2(z_1;z)=(|z_1|(2z-1))^2-3,$$
\begin{equation}\label{rel4}
A_3=p_3(z_1;z)=\frac{1}{2}|z_1|(2z-1)[(|z_1|(2z-1))^2-5],\;\
A_4=p_4(z_1;z)=\frac{1}{6}[(|z_1|(2z-1))^4+3(|z_1|(2z-1))^2 ].
\end{equation}

Using (\ref{honeig}) and (\ref{spechony}), the spectral distribution
is obtained as
$$\mu(z_1,\bar{{z_1}})=\frac{1}{9}\{\delta(z_1-3)\delta(\bar{z_1}-3)+2\delta(z_1-(2+\omega))\delta(\bar{z_1}-(2+{\omega}^2))+2\delta(z_1-(2+{\omega}^2))\delta(\bar{z_1}-(2+\omega))$$
\begin{equation}
+2\delta(z_1)\delta(\bar{z_1})+\delta(z_1-(2\omega+1))\delta(\bar{z_1}-(2{\omega}^2+1))
+\delta(z_1-(2{\omega}^2+1))\delta(\bar{z_1}-(2\omega+1))\}.
\end{equation}

 Using (\ref{pfh}), the probability  amplitude of observing the walk at starting site at
time $t$ is calculated as
\begin{equation}
 P_0(t)=\frac{1}{9}(\cos3t+6\cos \sqrt{3}t+2).
\end{equation}

Other probability amplitudes can be calculated using (\ref{amph})
and (\ref{rel4}).

\subsubsection{CTQW on honeycomb lattice}
 In the limit of the large $m$, the eigenvalues $z_{kl}$ reduce to $z_{kl}=1+e^{-ix_1}+e^{ix_2}$
where $x_1=\lim_{k,m\rightarrow\infty}\frac{2\pi k}{m}$ and
$x_2=\lim_{l,m\rightarrow\infty}\frac{2\pi l}{m}$. Therefore, the
continuous spectral distribution is
$$
\mu(z_1,\bar{{z_1}})=\int_{0}^{2\pi}\int_{0}^{2\pi}dx_1dx_2\delta(z_1-(1+e^{-ix_1}+e^{ix_2}))\delta(\bar{z_1}-(1+e^{ix_1}+e^{-ix_2}))=$$
\begin{equation}
\frac{1}{4{\pi}^2\sqrt{3-z_1^2{\bar{z_1}}^2-(z_1^2+{\bar{z_1}}^2)+2z_1\bar{z_1}(z_1+\bar{z_1})-2(z_1+\bar{z_1})}}.
\end{equation}

Therefore, in the limit of the large $m$, the probability amplitude
of observing the walk at level $k$ is given by
\begin{equation}
P_k(t)=\int_{0}^{2\pi}\int_{0}^{2\pi}dx_1dx_2(e^{i\sqrt{3+2(\cos
x_1+\cos x_2+\cos(x_1+x_2))}t}p_k(x_1,x_2;0)+ e^{-i\sqrt{3+2(\cos
x_1+\cos x_2+\cos(x_1+x_2))}t}p_k(x_1,x_2;1)).
\end{equation}
In particular, for the probability amplitude $P_0(t)$ we have
\begin{equation}
P_0(t)=\int\int
\cos(|z_1|t)\mu(z_1,\bar{{z_1}})d{z_1}d\bar{z_1}=\int_{0}^{2\pi}\int_{0}^{2\pi}
\cos \sqrt{3+2(\cos x_1+\cos x_2+\cos (x_1+x_2))}dx_1dx_2
\end{equation}
\subsubsection{asymptotic behavior}
Similar to the finite hexagonal lattice, we investigate the
asymptotic behavior of quantum walk at large time $t$ using
stationary phase approximation method. In this case we deal with the
integral
\begin {equation}\label{int6}
I_0(t)=\int_{0}^{2\pi}\int_{0}^{2\pi} e^{-i(\sqrt{3+2(\cos x_1+\cos
x_2+\cos (x_1+x_2))})t}dx_1dx_2+\int_{0}^{2\pi}\int_{0}^{2\pi}
e^{+i(\sqrt{3+2(\cos x_1+\cos x_2+\cos (x_1+x_2))})t}dx_1dx_2,
\end{equation}
as probability amplitude $P_0(t)$, so, by using the stationary phase
method we approximate the integrals for large time $t$. Here, we
have $g(x)=1$ and $f(x_1,x_2)=\sqrt{3+2(\cos x_1+\cos x_2+\cos
(x_1+x_2))}$. Then the asymptotic form of the probability amplitude
$P_0(t)$ is calculated as
\begin {equation}\label{asyh}
I_0(t)\simeq \frac{\pi}{t}(2\sqrt{3}sin3t+2cost).
\end{equation}
Now, we  compare the finite probability amplitude (\ref{pfh}) and
the continuous probability (\ref{asyh}) at large time $t$.
Therefore, we calculate numerically the difference between amplitude
of walk on honeycomb lattice and finite one, for large time $t$
\begin {equation}
\pi(m,t)=|I_0(t)-\frac{1}{m^2}\sum_{k,l=0}^{m-1}
\cos(\sqrt{3+2(\cos 2\pi k/m+cos 2\pi l/m)})t|.
\end{equation}
The result has been depicted in Fig.4. The figure shows that, the
difference $\pi(m,t)$ is limited to zero for $m\sim 60$ and $t\sim
700 $. Therefore, to study the behavior of asymptotic quantum walk
on finite honeycomb lattice, we can approximate it with infinite
one, and by using the stationary phase method, study the behavior
of asymptotic quantum walk.
\section{Generalization to $Z^{\otimes n}_m$}
Similar to the case $n=2$, we choose generating set
$P_{(10...0)}=\{(10...0),...,(0...01),(m-1...m-1)\}$ for
$Z^{\otimes n}_m$. Then, the orbits of Weyl group $S_{n+1}$ which
are indexed by $n$-tuple $m=(m_1,...,m_n)$, are given by
\begin{equation}
P_m=O((m_1,-m_2,-m_2-m_3,...,-m_2-m_3-...-m_n)).
\end{equation}
 By using (\ref{relation}) one can
obtain a translation invariant partition $R$ for $(Z^{\otimes
n}_m)^2$.  In the regular representation, the adjacency matrix of
underlying graph is written as
\begin{equation}\label{AzN}
A_{z_1}=S_1+...+S_n+(S_1...S_n)^{-1},
\end{equation}
with $S_i=I\otimes...\otimes \underbrace{S}_i\otimes
I\otimes...\otimes I$.  Clearly, the relations
 $R_m$ define an abelian  association scheme (not necessarily symmetric) on $Z^{\otimes
n}_m$. Moreover, the corresponding Bose-Mesner algebra is generated
by
\begin{equation}\label{generators}
A_{z_k}=\sum_{i_1<i_2<...<i_k}^{n+1}S_{i_1}S_{i_2}...S_{i_k} \;\ ,
\;\ k=1,...,n,
\end{equation}
where, $S_{n+1}=(S_1...S_n)^{-1}$. In the regular representation of
the group, for the corresponding adjacency matrices we have
\begin{equation}\label{adjacencyNN}
A_m=\sum_{g\in P_m}g\;\ .
\end{equation}

From (\ref{generators}) and (\ref{adjacencyNN}), it follows that the
adjacency matrices satisfy the following recursion relations
\begin{equation}\label{adj.recursion}
A_{z_k}A_{m}=\sum_{i_1<i_2<...<i_k}^{n+1}A_{m+v_{i_1}+...+v_{i_k}}\;\
, \;\ k=1,...,n,
\end{equation}
 where, $v_i$, for $i=1,...,n+1$, are $n$-dimensional vectors
 whose components are
\begin{equation}\label{Vi}
(v_i)_l=\delta_{l,i}-\delta_{l,i-1} \;\ , \;\ l=1,2,...,n.
\end{equation}
 In particular, $A_{m}=p_{m}(A_{z_1},...,A_{z_n})$, where $p_{m}$ is a polynomial
of degree $m_1+...+m_n$ with real coefficients. We refer to these
types of association schemes as $n$-variable $P$-polynomial
association schemes.

 Now, we symmetrize the graph to
obtain an undirected underlying graph as in the case of $n=2$. Then,
we will have for the adjacency matrix
\begin{equation}
 A=A_{z_1}+A_{z_n},
\end{equation}
where, $A_{z_1}$ is given by (\ref{AzN}) and $A_{z_n}=(A_{z_1})^t$.
The spectrum  of $A_{z_k}$ (indexed by $n$-tuple $l=(l_1,...,l_n)$),
is given by
\begin{equation}\label{EigN}
z^{(k)}_{l}=\sum_{i_1<i_2<...<i_k}^{n+1}\omega^{l_{i_1}+...+l_{i_k}},\;\;\
l_i\in\{0,1,...,m-1\},
\end{equation}
where, $\omega=\exp^{2\pi i/m}$ and $l_{n+1}=-(l_1+...+l_n)$. From
(\ref{EigN}), on can see that the spectrum of $A_{z_n}$ is complex
conjugate of the spectrum of $A_{z_1}$. Therefore, the eigenvalues
of $A$ are given by
\begin{equation}
\lambda_{l_1,...,l_n}=2(\cos2\pi l_1/m+...+\cos2\pi l_n/m+\cos2\pi
(l_1+...+l_n)/m).
\end{equation}
The spectral distribution associated with the generators is given by
 \begin{equation}\label{specN}
 \mu(z_1,...,z_n) = \frac{1}{m^n} \sum_{l_1,...,l_n}
  \delta(z_1-z^{(1)}_{l_1..l_n})\delta(z_2-z^{(2)}_{l_1..l_n})...\delta(z_n-z^{(n)}_{l_1..l_n}),
  \end{equation}
where $l_i \in \{0,1,...,m-1\}$, for $i=1,...,n$. Now using
(\ref{v4}) and spectral distribution (\ref{specN}), the amplitude of
observing the walk at level $i=(i_1,...,i_n)$ can be calculated as
\begin{equation}\label{pkN}
P_{i}(t)=\frac{1}{m^n}\sum_{l_1,...,l_n}e^{-2it(\cos{2\pi
{l_1}/m}+...+\cos{2\pi
{l_n}/m}+\cos{2\pi(l_1+...+l_n)/m})}p_{i}(z^{(1)}_{l_1..l_n},...,z^{(n)}_{l_1..l_n}),
\end{equation}
where, the polynomials $p_{i}(z_1,...,z_n)$ are given by the
following recursion relations
\begin{equation}\label{pkN}
z_kP_{m}=\sum_{i_1<i_2<...<i_k}^{n+1}P_{m+v_{i_1}+...+v_{i_k}}\;\
, k=1,...,n,
\end{equation}
where, $v_i$ for $i=1,...,n+1$ are defined by (\ref{Vi}). In
particular, the probability amplitude of observing the walk at
starting site at time $t$ is given by
\begin{equation}\label{p0N}
P_0(t):=\frac{1}{m^n}\sum_{k,l}e^{-2it(\cos{2\pi
{l_1}/m}+...+\cos{2\pi {l_n}/m}+\cos{2\pi(l_1+...+l_n)/m})}.
\end{equation}

  In the limit of large $m$, the eigenvalues $z_{l_1..l_n}$ reduce to
$z_{l_1..l_n}=e^{ix_1}+...+e^{ix_n}+e^{-i(x_1+...+x_n)}$ with
$x_i=\lim_{{l_i,m}\rightarrow \infty}2\pi {l_i}/m$, and the
spectral distribution reduces to continuous  constant measure
$\mu(x_1,..,x_n)=1/(2\pi)^n$. In fact, in the limit of large $m$,
the study of CTQW on finite symmetric graph constructed from
$Z^{\otimes n}_m$ as above, is equivalent to the study of walk on
the root lattice $A_n$, where the continuous form of probability
amplitude $P_i(t)$ is given by
\begin{equation}\label{contp0}
P_i(t)=\frac{1}{(2\pi)^n}
\int_{0}^{2\pi}...\int_{0}^{2\pi}e^{-2it(\cos x_1+...+\cos
x_n+\cos(x_1+...+x_n))}p_i(x_1,...,x_n)dx_1...dx_n,
\end{equation}
where, $A_i=p_i(x_1,...,x_n)$. In particular, the probability
amplitude of observing the walk at starting site at time $t$ is
given by
\begin{equation}\label{contp0}
P_0(t)=\frac{1}{(2\pi)^n}\int_{0}^{2\pi}...\int_{0}^{2\pi}e^{-2it(\cos
x_1+...+\cos x_n+\cos(x_1+...+x_n))}dx_1...dx_n,
\end{equation}
 where, the integral in (\ref{contp0})
can be approximated by employing stationary phase method at large
time $t$.
\section{conclusion}
Using the spectral distribution method, we investigated CTQW on root
lattice $A_n$ and honeycomb one, by constructing two types of of
association schemes and approximating the infinite lattices with
finite underlying graphs of constructed association schemes, for
large sizes of the graphs and large times. Although we focused
specifically on root lattice $A_n$ and honeycomb one, the underlying
goal was to develop general ideas that might then be applied to
other infinite lattices such as root lattices $B_n$, $C_n$ and etc.
also quasicrystals with certain symmetries. Apart from physical
results, we succeeded to obtain some interesting mathematical
results such as a generalization to the notion of $P$-polynomial
association scheme, where we expect that, the $n$-variable
$P$-polynomial association schemes posses the analogous properties
of $P$-polynomial association scheme and can be applied in coding
theory in order to construction of new codes. We hope that, studying
other infinite lattices leads us to other interesting mathematical
objects, perhaps new types of association schemes.

 \vspace{1cm}
\setcounter{section}{0}
 \setcounter{equation}{0}
 \renewcommand{\theequation}{A-\roman{equation}}
\textbf{Appendix : Suppersymmetric structure of $\Gamma_s$}
\\From the block form of $A$ in (\ref{honey}), we can see that the
constructed underlying graph of association scheme from two copies
of finite hexagonal lattice, has supersymmetric structure.
Following Ref.\cite{supersymetry}, we introduce the model of
supersymmetric algebra as follows

We define two operators $Q_+$ and $Q_-$ as
\begin{equation}
 Q_+=\left(\begin{array}{cc}
    O & O \\
    B & O \
  \end{array}\right) ; \;\;\ Q_-=\left(\begin{array}{cc}
    O & B^t \\
    O & O \
  \end{array}\right).
  \end{equation}
  Then we  define two hermitean charges $Q_1$, $Q_2$ and
  Hamiltonian $H$ as follows
  \begin{equation}
  Q_1= Q_+ + Q_- , \;\;\ Q_2= -i(Q_+ - Q_-),\;\;\ H=Q_1^2 =Q_2^2.
  \end{equation}
  With the above definitions, we get
  $$ Q_+^2=Q_-^2=0 ,\;\ H=\{Q_+,Q_-\}, \;\  [H,Q_{\pm}]=0 , \;\ [H,Q_{1,2}]=0,$$
  \begin{equation}
  \{Q_1,Q_2\}=0,\;\ \Rightarrow \;\ \{Q_i,Q_j\}=2H\delta_{ij}.
  \end{equation}
Therefore, $Q_+$, $Q_-$ and $H$ generate a closed supersymmetric
algebra.

For the association scheme derived from two copies of finite
hexagonal lattice in section $4$, we make the following
correspondence
\begin{equation}
Q_1=A=\left(\begin{array}{cc}
    O & B^t \\
    B & O \
  \end{array}\right),\;\;\ Q_2=\left(\begin{array}{cc}
    O & iB^t \\
    -iB & O \
  \end{array}\right),
  \end{equation}
  and therefore,
  \begin{equation}
  H=A^2=\left(\begin{array}{cc}
    B^tB & O \\
    O & BB^t \
  \end{array}\right); \;\;\
  Q_+=\left(\begin{array}{cc}
    O & O \\
    B & O \
  \end{array}\right),\;\;\ Q_-=\left(\begin{array}{cc}
    O & B^t \\
    O & O \
  \end{array}\right).
  \end{equation}
  In other words, the adjacency $A$ is our original Dirac
  operator. Now it can be checked that all the above (anti)
  commutation relations are fulfilled by our representation in
  the form of graph operators. In our special graph, $B$ and
$B^t$ commute with each other, so the Hamiltonian is of the form
$H=I \otimes B^tB$. Therefore,  the spectrum of $H$ is at least
twofold degenerate, i.e.,
\begin{equation}
H(f,g)^t=E(f,g)^t \;\ \Rightarrow \;\ B^tBf=Ef ,\;\;\  B^tBg=Eg.
\end{equation}
Hence, $(f,0)^t$ and $(0,g)^t$ are eigenvectors of $H$ to the same
eigenvalue.

 As  $H= Q_1^2= Q_2^2$  and  $\{Q_1,Q_2\}=0$, certain
combinations of the above eigenvectors yield common eigenvectors
of the pairs $H$, $Q_i$.\\ From the fact that $[B,B^t]=0$, we know
that $B$ and $B^t$ have common eigenvectors. If
\begin{equation}
B\textit{f}=\lambda \textit{f} ,\;\;\
B^t\textit{f}=\lambda^{'}\textit{f} , \end{equation}
 then
 \begin{equation}
BB^t\textit{f}=\lambda\lambda^{'}\textit{f}.
 \end{equation}
 As the
spectrum of $B$ is complex conjugate of the spectrum of $B^t$, we
have
\begin{equation}
 BB^t\textit{f}=|\lambda|^2\textit{f}.
 \end{equation}
 In other words, the eigenvector of $B$ with eigenvalue $\lambda$, is eigenvector of
Hamiltonian $H$, with eigenvalue $|\lambda|^2$ and degeneracy at
least $2$.

\newpage
{\bf Figure Captions} {\bf Figure-1: Shows root system corresponding
to $A_2$.}

{\bf Figure-2: Shows finite honeycomb lattice with generators $cI$,
 $cS_1^{-1}$ and $cS_2$.}

 {\bf Figure-3: Shows $\pi(m,t)$ for root lattice $A_2$
 as a function of $m$ (where, $m^2$ is the number of vertices of finite hexagonal lattice) at $t \sim 1000$,
  where the difference is almost negligible for $m\geq 50$.}

 {\bf Figure-4: Shows $\pi(m,t)$ for honeycomb lattice
 as a function of $m$ (where, $2m^2$ is the number of vertices of finite honeycomb lattice) at $t \sim 700$,
 where the difference is almost negligible for $n \geq 60$.}
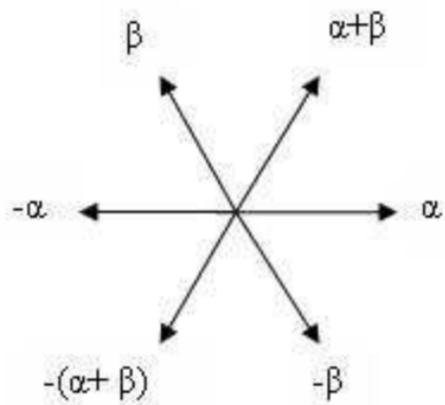
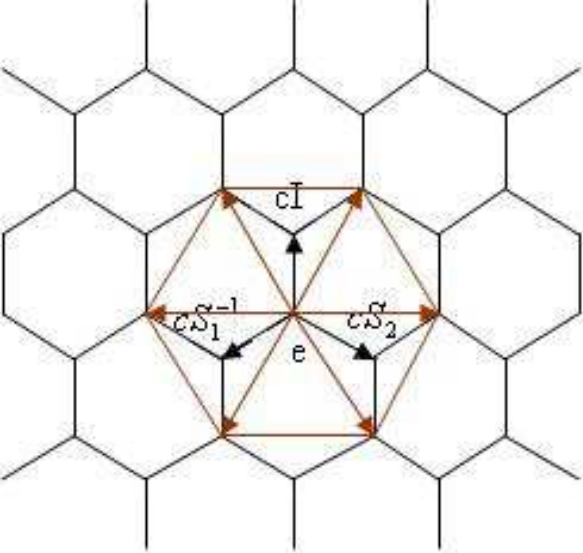
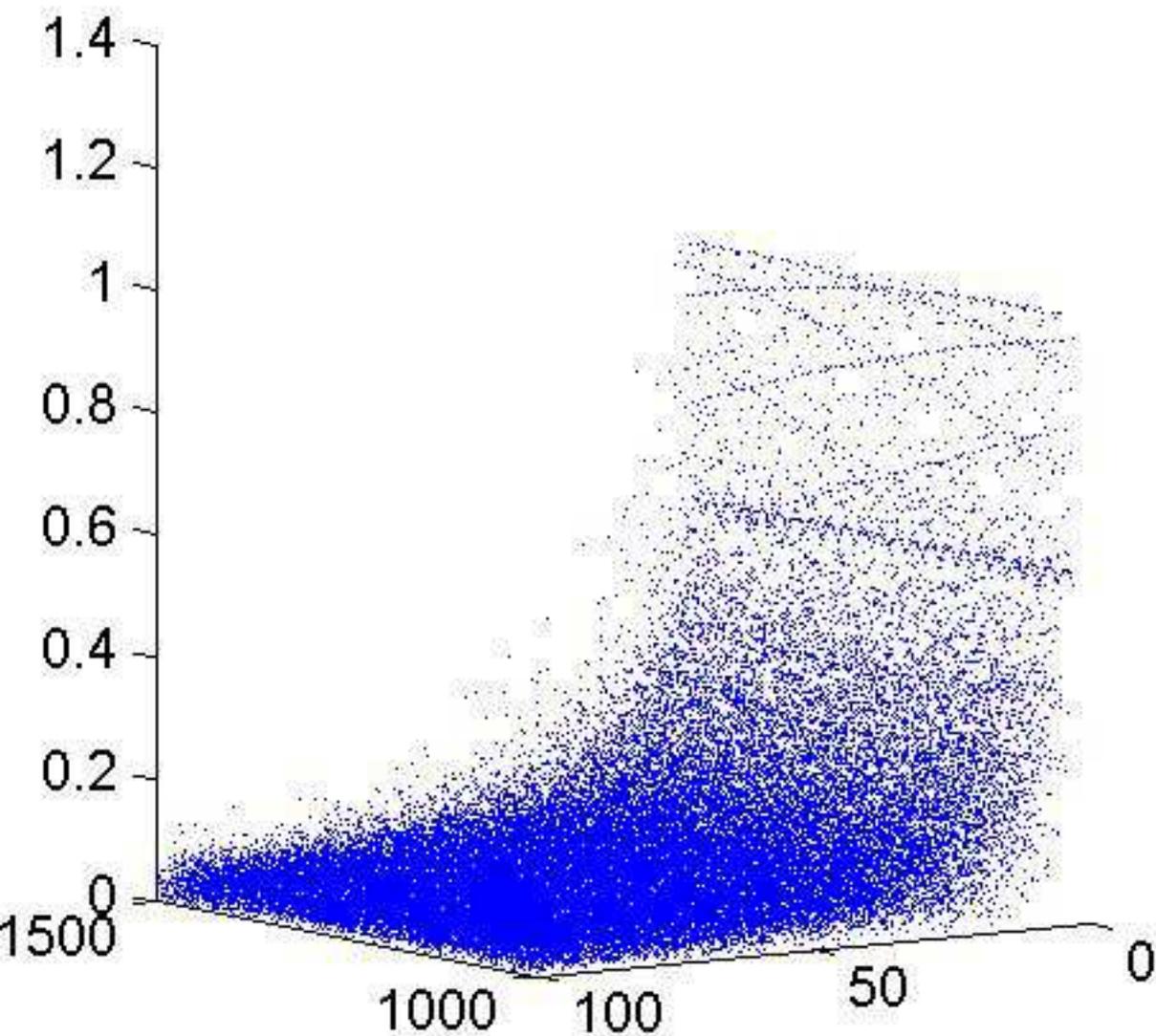
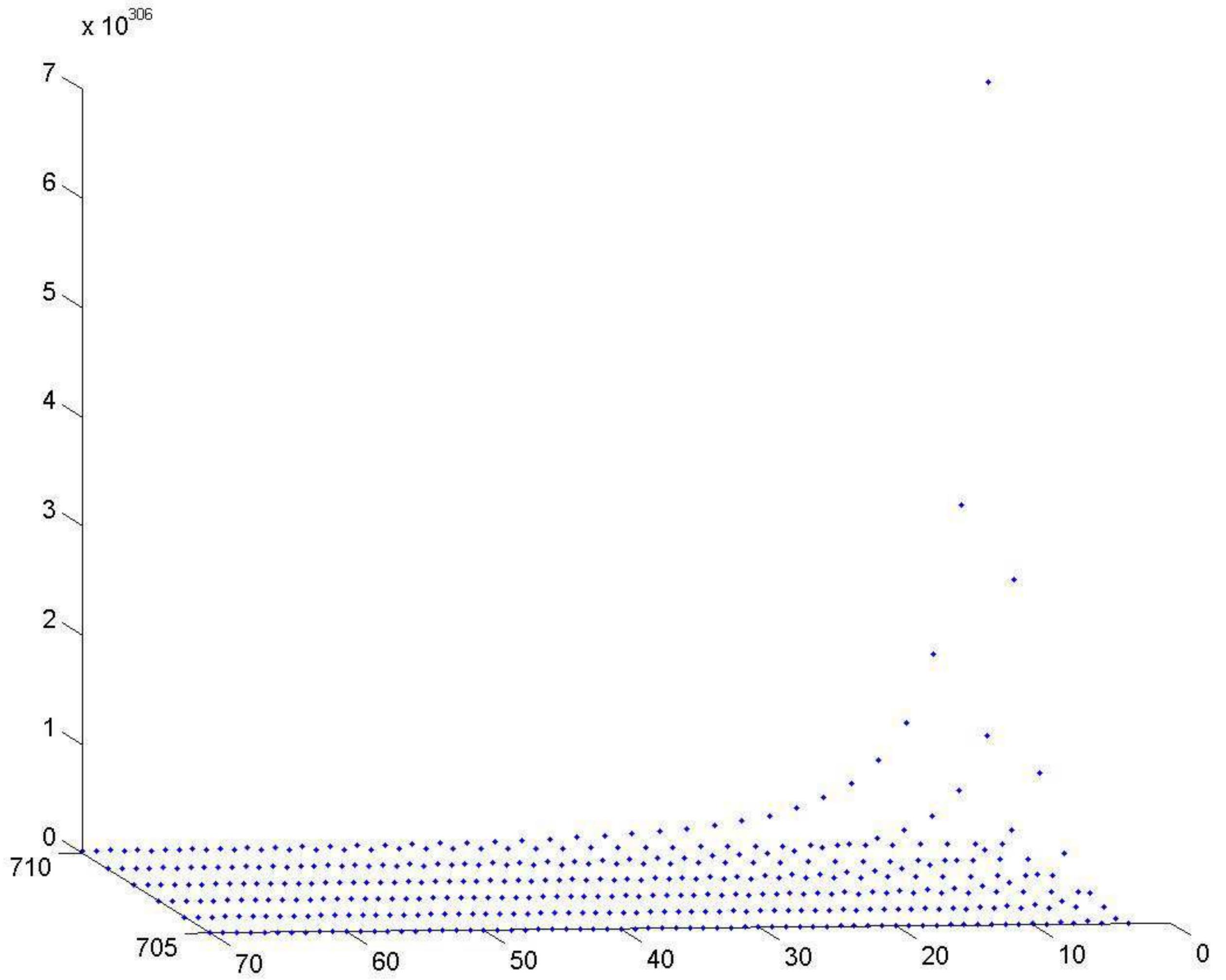
\end{document}